\documentclass{./aa-package/aa}
\usepackage[utf8]{inputenc}

\usepackage{graphicx}
\usepackage[breaklinks, colorlinks, citecolor=blue, linkcolor=blue]{hyperref}
\usepackage{tabularx}
\usepackage{xcolor}

\usepackage{etoolbox}
\robustify\bfseries

\usepackage{lscape}

\usepackage{multirow}

\usepackage{siunitx}
\DeclareSIUnit\pixel{pix}
\DeclareSIUnit\year{a}

\usepackage{placeins}

\newcommand{\CRX}{CRX}
\newcommand{\dLW}{dLW}
\newcommand{\CONTRAST}{CCF-contrast}
\newcommand{\FWHM}{CCF-FWHM}
\newcommand{\BISECTOR}{CCF-BIS}
\newcommand{\CaHtwo}{CaH~2}
\newcommand{\CaHthree}{CaH~3}
\newcommand{\CaIRT}{Ca\,{\sc ii}~IRT}
\newcommand{\CaIRTa}{Ca~IRT$_\text{a}$}
\newcommand{\CaIRTb}{Ca~IRT$_\text{b}$}
\newcommand{\CaIRTc}{Ca~IRT$_\text{c}$}
\newcommand{\Fe}{Fe~$\lambda$\SI{8689}{\angstrom}}
\newcommand{\Halpha}{H$\alpha$}
\newcommand{\He}{He~{\sc i} $\lambda$\SI{10830}{\angstrom}}
\newcommand{\HeDthree}{He\,{\sc i}~D$_3$}
\newcommand{\NaDdoublet}{Na\,{\sc i} doublet}
\newcommand{\NaDone}{Na~D$_1$}
\newcommand{\NaDtwo}{Na~D$_2$}
\newcommand{\Pabeta}{Pa$\beta$}
\newcommand{\TiOsevenzero}{TiO~$\lambda$\SI{7048}{\angstrom}}
\newcommand{\TiOeightfour}{TiO~$\lambda$\SI{8428}{\angstrom}}
\newcommand{\TiOeighteight}{TiO~$\lambda$\SI{8858}{\angstrom}}
\newcommand{\VOsevenfour}{VO~$\lambda$\SI{7434}{\angstrom}}
\newcommand{\VOsevenine}{VO~$\lambda$\SI{7940}{\angstrom}}
\newcommand{\WFB}{FeH WFB}

\makeatletter
\renewcommand*\aa@pageof{, page \thepage{} of \pageref*{LastPage}}
\makeatother

\let\orgautoref\autoref
\renewcommand{\autoref}
        {\def\equationautorefname{Eq.}%
         \def\figureautorefname{Fig.}%
         \def\sectionautorefname{Sect.}%
         \def\subsectionautorefname{Sect.}%
         \def\subsubsectionautorefname{Sect.}%
         \orgautoref}

\makeatletter
\def\instrefs#1{{\def\scsep{\def\scsep{,}}\@for\w:=#1\do{\scsep\ref{inst:\w}}}}
\renewcommand{\inst}[1]{\unskip$^{\instrefs{#1}}$}
\makeatother

\begin{document}

\title{The CARMENES search for exoplanets around M dwarfs}

\subtitle{Cluster analysis of signals from spectral activity indicators to search for shared periods}

\author{J.~Kemmer\inst{lsw} 
    \and M.~Lafarga\inst{warwick,ceh} 
    \and B.~Fuhrmeister\inst{ham} 
    \and Y.~Shan\inst{ceed,iag} 
    \and P.~Schöfer\inst{iaa,iag} 
    \and S.\,V.~Jeffers\inst{tls} 
    \and J.\,A.~Caballero\inst{cab} 
    \and A.~Quirrenbach\inst{lsw} 
    \and P.\,J.~Amado\inst{iaa} 
    \and A.~Reiners\inst{iag} 
    \and I.~Ribas\inst{ice,ieec} 
    \and V.\,J.\,S.~B\'ejar\inst{iac,ull} 
    \and F.~Del Sordo\inst{ice,ieec} 
    \and A.\,P.~Hatzes\inst{tls} 
    \and Th.~Henning\inst{mpia} 
    \and I.~Hermelo\inst{caha} 
    \and A.~Kaminski\inst{lsw} 
    \and D.~Montes\inst{ucm} 
    \and J.\,C.~Morales\inst{ice,ieec} 
    \and S.~Reffert\inst{lsw} 
}

\institute{
\label{inst:lsw}Landessternwarte, Zentrum für Astronomie der Universität Heidelberg, Königstuhl 12, 69117 Heidelberg, Germany 
\and \label{inst:warwick}Department of Physics, University of Warwick, Gibbet Hill Road, Coventry CV4 7AL, United Kingdom
\and\label{inst:ceh}Centre for Exoplanets and Habitability, University of Warwick, Coventry, CV4 7AL, United Kingdom
\and \label{inst:ham}Hamburger Sternwarte, Gojenbergsweg 112, 21029 Hamburg, Germany
\and \label{inst:ceed}Centre for Earth Evolution and Dynamics, Department of Geosciences, Universitetet i Oslo, Sem S{\ae}lands vei 2b, 0315 Oslo, Norway
\and \label{inst:iag}Institut f\"ur Astrophysik, Georg-August-Universit\"at, Friedrich-Hund-Platz 1, 37077 G\"ottingen, Germany
\and \label{inst:iaa}Instituto de Astrof\'isica de Andaluc\'ia (IAA-CSIC), Glorieta de la Astronom\'ia s/n, 18008 Granada, Spain
\and \label{inst:tls}Th\"uringer Landessternwarte Tautenburg, Sternwarte 5, 07778 Tautenburg, Germany
\and \label{inst:cab}Centro de Astrobiolog\'ia (CAB, CSIC-INTA), Camino Bajo del Castillo s/n, Campus ESAC, 28692 Villanueva de la Ca\~nada, Madrid, Spain
\and \label{inst:ice}Institut de Ci\`encies de l'Espai (ICE, CSIC), c/ de Can Magrans s/n, Campus UAB, 08193 Cerdanyola del Vall\`es, Spain
\and \label{inst:ieec}Institut d'Estudis Espacials de Catalunya (IEEC), c/ Gran Capit\`a 2-4, 08034 Barcelona, Spain
\and \label{inst:iac}Instituto de Astrof\'isica de Canarias (IAC), 38205 La Laguna, Tenerife, Spain
\and \label{inst:ull}Departamento de Astrof\'isica, Universidad de La Laguna, 38206 La Laguna, Tenerife, Spain
\and \label{inst:mpia}Max-Planck-Institut f\"{u}r Astronomie, K\"{o}nigstuhl  17, 69117 Heidelberg, Germany
\and \label{inst:caha}Centro Astron\'onomico Hispano en Andaluc\'ia, Observatorio de Calar Alto, Sierra de los Filabres, 04550 G\'ergal, Spain
\and \label{inst:ucm}Departamento de F\'isica de la Tierra y Astrof\'isica \& IPARCOS-UCM (Instituto de F\'isica de Part\'iculas y del Cosmos de la UCM), Facultad de Ciencias F\'isicas, Universidad Complutense de Madrid, 28040 Madrid, Spain
}

\date{Received 30 May 2023 / Accepted 26 February 2025}

\abstract
{A multitude of spectral activity indicators are  routinely computed nowadays from the spectra generated as part of planet-hunting radial velocity surveys. Searching for shared periods among them can help to robustly identify astrophysical quantities of interest, such as the stellar rotation period. However, this identification can be complicated due to the fact that many different peaks occurring in the periodograms. This is especially true in the presence of aliasing and spurious signals caused by environmental influences affecting the instrument.}
{Our goal is to test a clustering algorithm to find signals with the same periodicity, (i.e. with the stellar rotation period) in the periodograms of a large number of activity indicators. On this basis, we have looked to evaluate the correlations between activity indicators and fundamental stellar parameters.}
{We used generalised Lomb-Scargle periodograms to find periodic signals in 24 activity indicators, spanning the VIS and NIR channels of the CARMENES spectrograph.
Common periods were subsequently determined by a machine learning algorithm for density-based spatial clustering of applications with noise ({\tt DBSCAN}).}
{The clustering analysis of the signals apparent in the spectral activity indicators is a powerful tool for the detection of stellar rotation periods.
It is straightforward to implement and can be easily automated, so that large data sets can be analysed. For a sample of 136 stars, we were able to recover the stellar rotation period in a total of 59 cases, including 3 with a previously unknown rotation period. In addition, we analysed spurious signals frequently occurring at the period of one year and its integer fractions,  concluding that they are likely aliases of one underlying signal. Furthermore, we reproduced the results of several previous studies on the relationships between activity indicators and the stellar characteristics.
}
{}


\keywords{techniques: radial velocities -- stars: late-type -- stars: low-mass -- stars: activity -- stars: rotation
}

\maketitle
%

\section{Introduction}

The intrinsic variability of stars that possess an outer convective zone is one of the main challenges in detecting and characterising exoplanets orbiting cool stars with high-resolution spectroscopy. The presence of magnetic activity features on the stellar surface, such as cool spots, hot faculae, or surface granulation, distort the symmetrical shape of the absorption line profiles used to measure radial velocities (RVs). As a result,   a true Doppler shift could end up hidden due to an orbiting planet \citep[e.g.][]{Desort2007,Barnes2011,Liebing2021A&A...654A.168L,Jeffers2022} or end up recorded as a false-positive planet \citep[e.g.][among many examples]{Queloz2001,Desidera2004,Huelamo2008,Figueira2010,Hatzes2013,Rajpaul2016,Haywood2014,Santos2014,Robertson2015}. Stellar activity features such as starspots can have lifetimes longer than the stellar rotation period and their impact on the RVs could be coincident with the stellar rotation period \citep[e.g.][]{Boisse2011, Haywood2014, Rajpaul2015, Stock2020a,Stock2020b, Kossakowski2022}. Stars also display long-term magnetic cycles that change their overall activity level \citep[e.g.][]{GomesdaSilva2011,Brandenburg2017,DiezAlonso2019,Terrien2022,Jeffers2023SSRv..219...54J,Fuhrmeister2023}.

Several indicators of stellar activity are often used to estimate stellar rotation periods and long-term cycles, as well as to disentangle stellar-induced RVs signals from those induced by true planetary companions. These include photometric time series that measure brightness variations on the stellar surface, as well as time series of spectroscopic indicators derived from the same data used to measure RVs \citep[e.g.][]{Queloz2001,Boisse2009,Astudillo-Defru2017b,Kemmer2022}. Common spectroscopic activity indicators include measurements of excess emission flux in the core of chromospheric lines \citep[e.g.][]{Wilson1968,Baliunas1995,West2004,Houdebine2009,Jeffers2018,Schofer2019}, the strength of photospheric bands \citep[e.g.][]{Berdyugina2002,Lepine2007,Schofer2019}, indicators tracing average shape changes in absorption lines such as those derived from the cross-correlation function (CCF) or template matching approaches \citep[e.g.][]{Queloz2001,GomesdaSilva2012,Lafarga2020, Barnes2024MNRAS.534.1257B}, or wavelength-dependent RV changes traced by the chromatic index \citep[CRX;][]{Zechmeister2018,Tal-Or2018,Jeffers2022,Jeffers2024}. Furthermore, magnetic field proxies derived from polarimetric measurements can also be used as activity indicators \citep[e.g.][]{Fouque2023}.

Several works have studied the temporal behaviour of activity indicators in cool stars, and how they relate to each other as well as to the RV measurements \citep[e.g.][]{Martinez-Arnaiz2010,Lovis2011,Robertson2013,SuarezMascareno2015,SuarezMascareno2017,SuarezMascareno2018,Mignon2023}. However, we still do not have a clear picture of how different indicators behave with respect to different types of stars. In this work, we focus on M dwarf stars observed within the CARMENES survey \citep{Quirrenbach2014,Ribas2023}, a sample that spans all M dwarf sub-spectral types with varying average activity levels. Alongside the search for exoplanets, studying the activity of M dwarfs has always been a central interest of the CARMENES survey. A number of studies have previously investigated the sensitivity to activity of spectroscopic indicators in the wavelength range of CARMENES. \cite{Tal-Or2018} found correlations between RV and CRX for about a third of stars in a sample of $\sim$30 stars showing large RV scatter. \cite{Schofer2019} studied the temporal variability of chromospheric indicators and photospheric absorption band indices in the CARMENES sample (composed of 331 M dwarfs at the time). Out of 133 stars with known stellar rotation period longer than 1 day, this study identified 15 stars with a significant periodic signal related to the rotation period in more than two indicators.
The authors found that the indicators most likely to vary with the rotation period were the \TiOsevenzero{} and \TiOeightfour{} photospheric band indices, along with the \Halpha{} and the \CaIRTb{} indices. In a subsequent work, 
\cite{Schofer2022} focussed on an in-depth study of four stars with different activity levels and sub-spectral types. The authors found that RVs and photospheric indicators (especially the \TiOsevenzero{} index) clearly vary with the stellar rotation period, while chromospheric indicators only show clear signals for low activity levels. Moreover, the authors also observed changes over time in the dominant harmonic, that is, episodes in which indicators vary with the rotation period and other episodes where they vary with half the rotation period.
\cite{Lafarga2021} studied a sub-sample of 98 stars and found that indicators behave differently depending on the mass and activity level of the stars. The authors found activity-related periodic signals in 56 stars of the sample in various indicators.
Stars with low activity levels tend to show signals in chromospheric indicators, while indicators such as CRX and CCF bisector inverse slope (\BISECTOR{}) show signals in the high-activity regime. Most stars show a signal related to the rotation period in the RVs and about half of them showed signals in the differential line width (\dLW{}) and CCF full width at half maximum (\FWHM{}), especially fully convective M dwarfs. 

\cite{Fuhrmeister2019a} presented a comparison of different methods to search for long-term periodic signals or cycles in the time-series of CARMENES activity indicators. They investigated the variability of the \Halpha{}, \NaDdoublet{}, and \CaIRT{} indices for \num{16} stars and found that Gaussian process regression and the classical periodogram analysis provide the best results.

However, all these studies were focussed on individual lines and partial aspects of the properties and behaviour of the activity indicators, as such detailed evaluations are not feasible on a large basis in a meaningful way. This is mainly due to the large number of spectroscopic activity indicators, which can have different sensitivities to rotational periods and can be related to stellar properties in complex ways. Furthermore, the interpretation of the multitude of signals in the time series of the indicators, including the need to filter out contamination by non-astrophysical signals, is non-trivial. Meanwhile, machine learning techniques provide promising ways to tease out meaningful signals from large collections of data. Clustering algorithms, for instance, are often used to find groups with similar properties. A prominent example is the detection of star clusters in the data of astrometric satellite missions \citep[e.g.][]{Caballero2008, Castro-Ginard2018, Hunt2021}. 
In this work, we used the density-based spatial clustering of applications with noise \citep[{\tt DBSCAN};][]{Ester1996} to detect signals that coincidentally occur in the periodograms of time series of several stellar activity indicators. In doing so, we took advantage of the fact that although each individual indicator may hold an ambiguous significance for a particular star, analysing a set of indicators can reveal patterns that help identify activity related signals, such as the stellar rotation period or long-term trends, as well as common spurious signals. The analysis is based on an updated data set from the CARMENES survey, which includes the full set of routinely computed activity indicators for all stars in the survey.

The paper is structured as follows. In \autoref{sec:data} we give an overview on the stellar sample and spectral activity indicators that we analyse. Our methods and the procedure to determine the clusters of signals in the indicators are described in \autoref{sec:analysis}. The results from the application of the clustering algorithm for the robust detection of stellar rotation periods and activity cycles are presented and discussed in 
Sects.~\ref{sec:results} and~\ref{sec:discussion}.  
Finally, we highlight our conclusions in \autoref{sec:conclusion}.


\section{Data}
\label{sec:data}

\subsection{Stellar sample}
\label{subsec:stellar_sample}
\begin{figure*}
    \centering
    \includegraphics[width=0.90\textwidth]{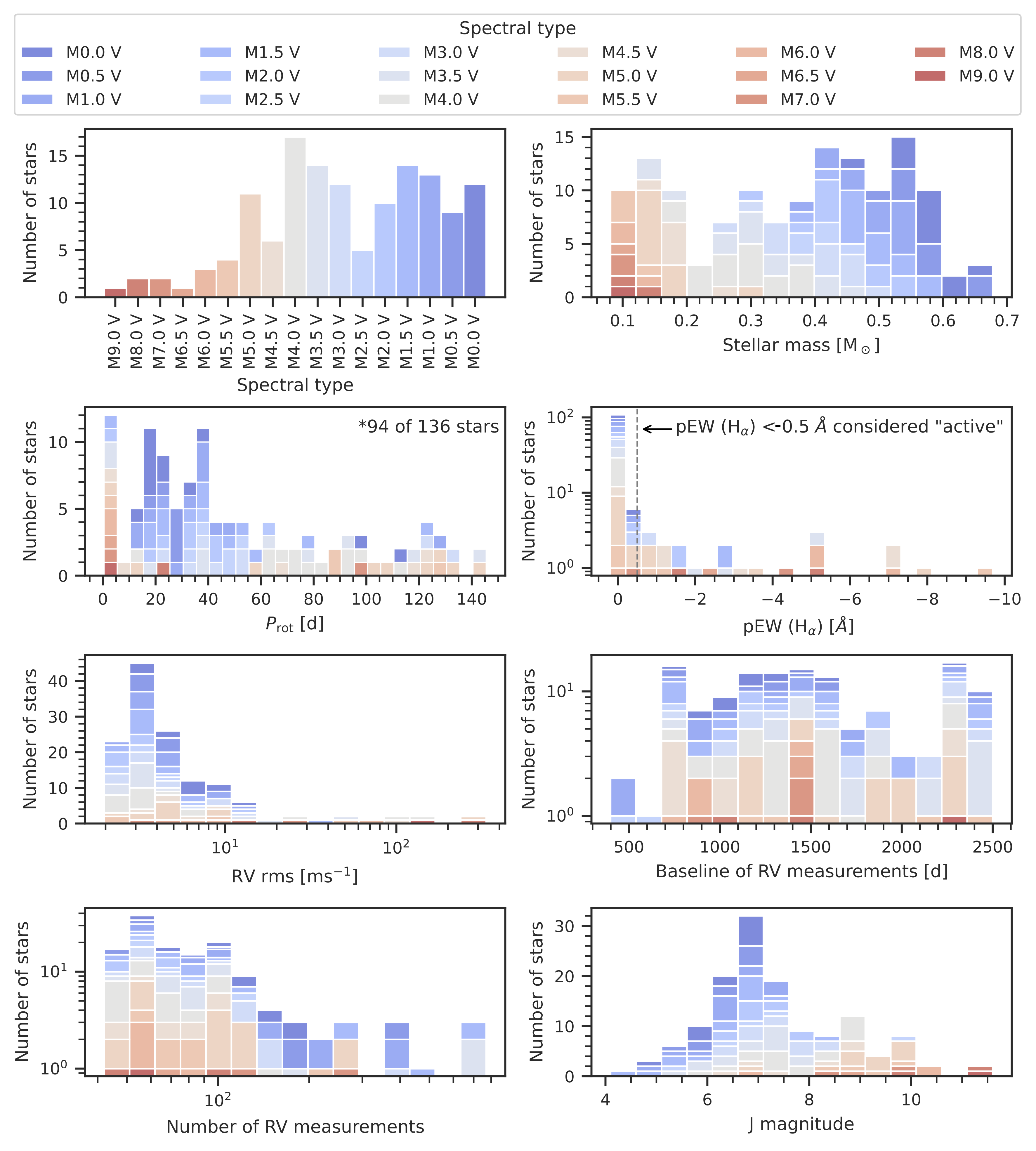}
    \caption{Overview on the properties of the stellar sample.}
    \label{fig:stellar_sample}
\end{figure*}

Our analysis was based on 362 stars from the CARMENES sample of M dwarfs \citep{Quirrenbach2016,Reiners2018b,Ribas2023}. We used observations taken between January 2016 and November 2022 as part of the guaranteed time observations and the legacy project of CARMENES. To ensure an adequate sampling, we selected for our analysis only M dwarfs from the CARMENES sample that had a minimum number of 40 observations over a period of at least one year. Furthermore, we excluded data with large gaps of more than \SI{40}{\percent} of the time of the observations, as these gaps can lead to severe aliasing. As in \cite{Lafarga2021} and \cite{Fuhrmeister2023}, we also removed all binary stars from the sample that were identified by \cite{Baroch2018} and \cite{Baroch2021}. The resulting final sample consists of more than \num{14000} spectra from 136 stars, out of which 97 have a known stellar rotation period \citep{Shan2024}. We show an overview of various key properties of the sample in \autoref{fig:stellar_sample}.

\subsection{Spectral activity indicators}

The advantage of using CARMENES spectra to analyse stellar activity indicators is its broad wavelength range. There are 24 activity indicators covered by the VIS and NIR channels of CARMENES as listed in \autoref{tab:overview_activity_indicators}. As outlined in the introduction, their sensitivity to stellar activity of M dwarfs and their predictive power for the stellar rotation period have already been investigated in detail in several studies of the CARMENES consortium (see \autoref{tab:overview_activity_indicators} for a list of relevant references). Generally, the indicators can be divided into three different categories:
($i$) chromospheric line indicators: the largest category, which includes \Halpha{}, for which line indices \citep{Zechmeister2018} or (pseudo-)equivalent-widths \citep[pEW,][]{Schofer2019} can be determined for our M-dwarf stars; ($ii$) photospheric absorption band indices: For broader photospheric lines, such as the TiO lines, no (pseudo-)continuum can be determined, so we used band indices instead \citep{Schofer2019};
and ($iii$) broad spectrum diagnostics: In addition, there are a number of indicators that are diagnostic to the whole spectrum (hereafter diagnostic indicators). These include the RV measurements themselves, as well as the CCF parameters \citep{Lafarga2020} and \dLW \citep{Zechmeister2018}, which are sensitive to changes in the shape of the spectral lines caused by activity. Furthermore, there is \CRX \citep{Zechmeister2018}, which expresses the wavelength dependence of the measured RVs.

We present an overview of all the activity indicators analysed in this work and the references for the methods used to calculate them in \autoref{tab:overview_activity_indicators}. We note that there are different definitions of the line indices in the presented literature. However, these differences do not have a major impact on our results since we are only interested in periodic signals in each of them individually and do not compare their absolute values \citep{Fuhrmeister2019a}.


\section{Analysis}
\label{sec:analysis}

\subsection{Periodogram analysis}
\label{subsec:periodogram_analysis}

\subsubsection{Method}
For our periodogram analysis, we used the \texttt{astropy} \citep{AstropyCollaboration2018} implementation of the generalised Lomb-Scargle (GLS) periodogram normalised to unity \citep{Zechmeister2009} with an oversampling factor of ten to properly resolve the peaks \citep{VanderPlas2018}. Our periodogram analysis considered formal uncertainties of the activity indicators, thereby accounting for differences in quality of the spectra used to calculate them (i.e. signal-to-noise ratio of the spectra, influences from flares, etc.). The false alarm probabilities (FAPs) were calculated using the analytic expression of \cite{Baluev2008}.

There are two different approaches to identify the signals from the periodograms. The naive method is to create the periodogram and note down the periods of all visible peaks above a given FAP threshold. However, because the GLS periodogram only considers one period at a time, dominant signals can completely suppress lower amplitude signals that may be missed in this way. A more sophisticated approach is to perform an iterative pre-whitening of the data, as is usually done when searching for planets in RV data. We performed this pre-whitening by generating the periodogram and finding the period with the lowest FAP. Next, the sinusoidal model corresponding to the signal was subtracted from the data and a new periodogram was generated from the residuals. The process was then repeated until no more signals above the specified FAP threshold were detected.

An effect of the pre-whitening method is that it also removes alias peaks occurring in the periodogram. This removal has pros and cons for the clustering algorithm (\autoref{subsec:clustering}). The advantage is that it significantly reduces false-positive clusters and crowding at the alias periods of the signals present in the data. On the other hand, the true underlying period creating the aliases does not always have the highest power in the periodogram \citep{Dawson2010,VanderPlas2018}. If during the pre-whitening an alias of the true period is removed (instead of removing the true period), this leads to a loss of data points in the cluster corresponding to the correct period, which may cause the clustering algorithm to miss it.

In the top two panels of \autoref{fig:clustering_teegarden}, we show as an example the results from the clustering algorithm obtained using the two different approaches to determine the signals from the periodograms. 
Because the aliases have been removed by the pre-whitening method, significantly fewer clusters were found in that case. The width of the individual clusters in the pre-whitening approach is significantly narrower because the close aliases caused by long-term sampling frequencies in the window function of the data are diminished. For our clustering analysis, we used both approaches: the naive method and the pre-whitening method. We compared their performance regarding the success in detecting the stellar rotation periods.

\subsubsection{Signal search}

Following \cite{Lafarga2021}, we removed outliers and data points with large uncertainties using a sigma clipping with a threshold of \SI{3}{$\sigma$} in the absolute values and uncertainties before creating the periodograms. We performed this clipping to exclude observations during stellar flares and, in general, to exclude observations with a poor signal-to-noise ratio (S/N).

The power of long-period signals is often greater in their one-day aliases than in the actual periods \citep{Dawson2010}. For this reason, we chose a lower bound of \SI{1.5}{\day} for the period range to minimise the impact of the one-day alias and to facilitate an automatic detection of the correct signals. We note that this cut can affect the signal recovery of fast-rotating stars, which tend to be either very young or late-M dwarfs. Very fast rotators comprise a very small fraction of our sample, that is, 4/136 of our targets have a confirmed $P_\mathrm{rot}<$\SI{1.5}{\day} and they are all late-M dwarfs \citep[][see \autoref{tab:rotation_periods} for details]{Shan2024}. We note that fast rotator candidates may be identified via spectroscopic $v\sin i$, since $P_\mathrm{rot}=$\SI{1.5}{\day} corresponds to an equatorial surface rotation velocity of between 3 and 20~km\,s$^{-1}$ (for late to early M dwarfs).
We further note that since very short signals are severely under-sampled with spectroscopic monitoring, methods based on spectroscopic activity indicators are inefficient at detecting very short signal compared to using photometric data.
Since young stars tend to be fast rotators, our cut at \SI{1.5}{\day} could also be biasing our period recovery of young stars. We note that it is hard to define a clear cut in period that separates young and old M dwarfs \citep[see e.g.][where the authors find young stars with periods above \SI{2}{\day}, but old, very late M dwarfs with periods below \SI{1}{\day}]{Cortes-Contreras2024}. According to \citet{Cortes-Contreras2024}, none of the four short-period stars mentioned above have been classified as young. Therefore, it remains unclear how our lower period boundary can bias our results regarding the youth of our targets.

By limiting the upper range of the period grid, it is (in principle) possible to improve the power of signals with a potential origin from stellar rotation, as we would expect those to be generally lower than \SI{200}{\day} \citep[e.g.][]{Newton2016,Newton2018,Shan2024}. However, in our analysis we were also interested in potential long-term activity cycles, which can, as mentioned above, significantly affect the power of shorter periods. Therefore, we adopted the standard in the GLS implementation from \texttt{astropy}, which is two times the timespan of our observations (hereafter baseline) multiplied by the oversampling factor.

An important parameter for the signal search is the FAP threshold, which is usually set between \SI{1}{\percent} and \SI{0.1}{\percent} in periodogram analysis. Since the clustering algorithm combines the signals from periodograms of multiple indicators, we took advantage of the fact that the FAP for a cluster of signals to occur at a specific period in multiple periodograms is much lower than the FAP of a single periodogram \citep[see Sect. 7.10.1 in][]{Hatzes2019}. We found that a FAP threshold of $\text{FAP}_\text{max} \leq \SI{80}{\percent}$ in the individual periodograms provides a good balance between the number of detected spurious signals and the recovery rate of the stellar rotation period, as shown in \autoref{subsec:rotation_periods}. In cases where the periodograms suffered from severe aliasing, a FAP threshold as the only limit in the signal search led to tens or even hundreds of spurious signals detected. Therefore, we limited the search to no more than 25 detected signals per activity indicator. In this way, summed up over all activity indicators, we detected an average of eight periodic signals per target with the naive approach and three periodic signals through pre-whitening.

\subsection{Clustering analysis}
\label{subsec:clustering}
\begin{figure*}[p]
    \centering
    \includegraphics[width=0.95\textwidth]{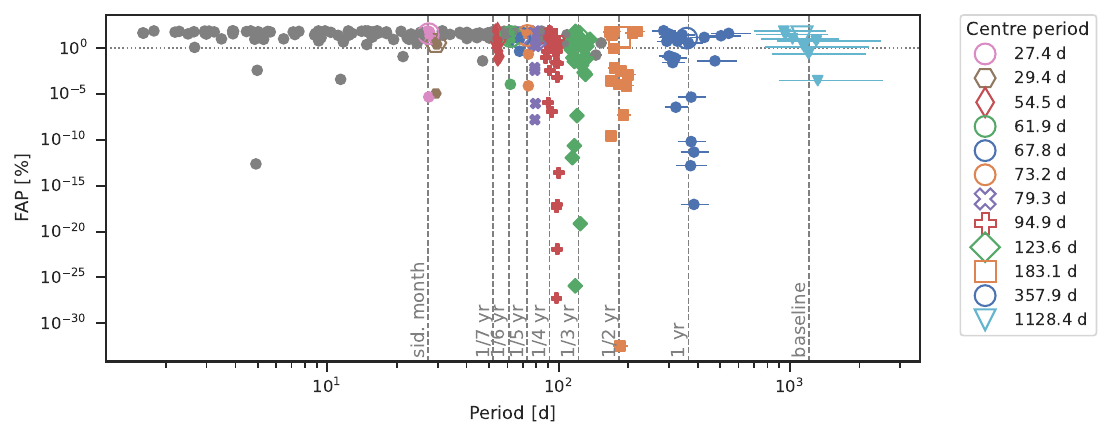}\\
    \includegraphics[width=0.95\textwidth]{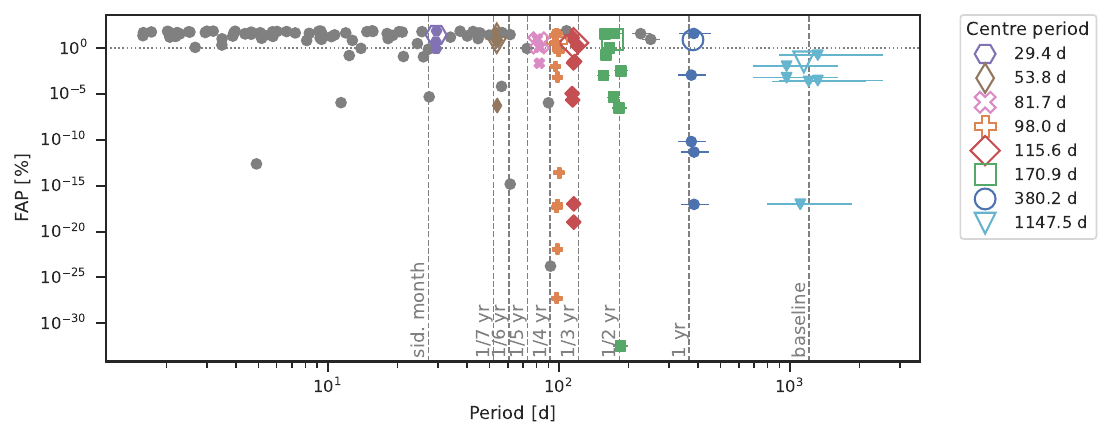}\\
    \includegraphics[width=0.95\textwidth]{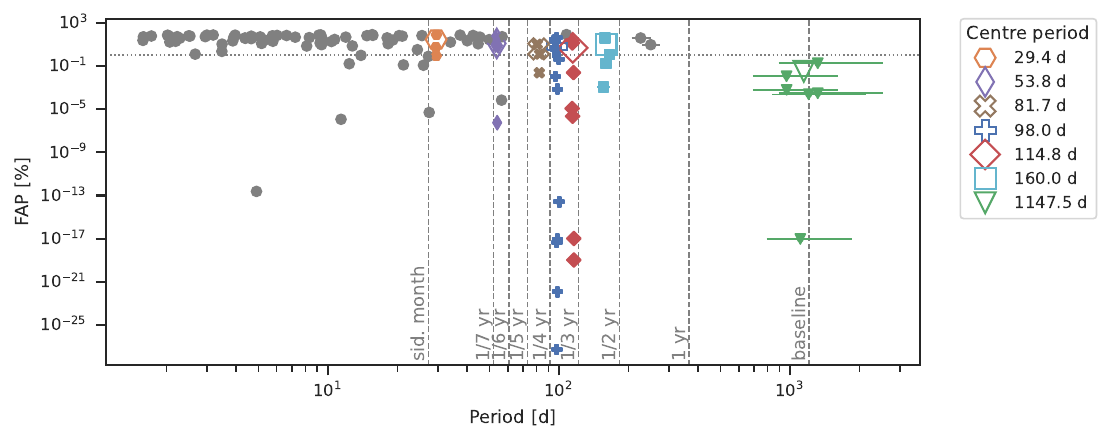}
    \caption{Results of the clustering algorithm for Teegarden's Star. \textit{Top}: Using the naive approach. \textit{Middle}: Using the pre-whitening approach. \textit{Bottom}: Using pre-whitening and excluding periods close to the harmonics of one year. The coloured filled symbols show the individual data points of the detected clusters, with the error bars corresponding to the peak width in the GLS, and the open symbols yield the mean values of the clusters. Grey dots denote data points that were identified as noise by the clustering algorithm. 
  The colour mapping changes between the plots to emphasise that although there is some overlap, the clusters (and thus technically also the mean periods) that appear  change between the different methods. Clusters that are consistent between all plots within the GLS resolution are always denoted by the same non-circular markers. The FAP of \SI{1}{\percent} is marked by the grey horizontal dotted line and the grey vertical dashed lines show different periods of interest. The baseline indicates the timespan of the observations.}
    \label{fig:clustering_teegarden}
\end{figure*}

\subsubsection{Method}

To investigate the occurrence of clustering in the periodicities detected in all \num{24} activity indicators listed in \autoref{tab:overview_activity_indicators}, we used the {\tt DBSCAN} algorithm implemented in \texttt{scikit-learn} \citep{Pedregosa2011}. Compared to other popular clustering algorithms, it has the advantage that it can also take noise into account (i.e. it is robust to outliers that do not belong to any cluster). This feature is particularly important for our analysis, as many spurious signals emerge from the periodogram analysis due to our low FAP thresholds. The {\tt DBSCAN} algorithm is defined by two parameters: the distance parameter $\epsilon$, which describes the distance between two data points that are to be considered being in a `neighbourhood' relation and min$_\text{data points}$, which is the minimum number of data points required for such a neighbourhood to be considered a cluster. In our case the data points are the measured signals from the periodogram analysis and the distance is the difference in their periods or frequencies. 
Tuning the $\epsilon$ parameter changes the resolution of the clustering algorithm. If it is too large, clusters may be merged, leading to incorrect period estimates. On the other hand, if the $\epsilon$ parameter is too small, the clusters may be fragmented, resulting in the loss of data points that are incorrectly marked as outliers.  The minimum number of data points basically determines the density of the clusters. However, as we are dealing with a relatively small number of data points, min$_\text{data points}$ is (in our context) more a measure of the reliability of a cluster. The more data points in a cluster, the more likely it is that there is a physical reason for the signals to occur at the same period rather than by chance.

\subsubsection{Cluster search}
\label{subsubsec:cluster_search}

We used a one-dimensional (1D) input for the {\tt DBSCAN} method, which represents the frequencies\footnote{So far, we have always referred to the periods, as they are more intuitive. For a better understanding of the actual implementation of the clustering algorithm, however, it is necessary to switch to the frequency domain in \autoref{subsubsec:cluster_search}, as this is the natural unit of the periodograms.} of the signals that we determined for each target in the periodogram analysis of the activity indicators. To be consistent with the signal search, we set the $\epsilon$ parameter of the {\tt DBSCAN} algorithm to half of the peak width in the GLS periodograms ($ \epsilon =1/2\times\text{baseline}$ in frequency space). In the classical periodogram analysis, a significant signal in the periodogram of both RVs and a single activity indicator is often reason enough to attribute said signal to stellar activity. For this reason, we kept the required minimum number of data points per cluster small and set it to 3. As a result, a cluster in our analysis corresponds to the agglomeration 
of at least three signals from the GLS periodograms of different activity indicators, where the difference in frequency to the nearest neighbour is not greater than the resolution of the periodograms. Due to the low FAP threshold in the periodogram analysis, this definition resulted in many spurious clusters consisting only of data points with high FAP. Therefore, we introduced an additional criterion for our analysis, and only considered clusters in which at least one data point meets the classical requirements and has a FAP of less than \SI{1}{\percent}.

A notable first result was that the clustering of the activity indicators in most cases shows very strong signals around the period of one year and its integer fractions (see the upper panel in \autoref{fig:clustering_teegarden}). Using the pre-whitening method in the periodogram analysis can remove some of these signals, as they are created by aliasing (middle panel in \autoref{fig:clustering_teegarden}), but the often very large clusters around the one and half year period are largely unaffected by this pre-whitening. In order to facilitate an automated evaluation to find signals with an activity origin as presented in Sects.~\ref{subsec:rotation_periods} and~\ref{subsec:other_activity},
we therefore removed from the analysis all peaks that were consistent with integer fractions (up to 1/6) of one year within the uncertainties of the GLS before applying the clustering algorithm. An exemplary result for Teegarden's star is shown in the lower panel of \autoref{fig:clustering_teegarden}.


\section{Results}
\label{sec:results}

\subsection{Detecting the stellar rotation period}
\label{subsec:rotation_periods}

\subsubsection{An example: Teegarden's Star}

The search for the stellar rotation period is the most obvious application of the clustering algorithm of the activity indicators. As an example, we show the results from Teegarden's Star in \autoref{fig:clustering_teegarden}. Teegarden's Star is host to at least three  (approximately )Earth-mass planets, with the inner two in its liquid water habitable zone \citep{Zechmeister2019, Dreizler2024}.
The very low-mass star shows additional RV signals that might be related to stellar activity.  
It has a spectroscopically determined period of \SI{\sim100}\,d, as seen in several spectroscopic activity indicators \citep[96.2, 97.6, 99.6\,d, depending on the indicator and dataset; see][]{Lafarga2021, Terrien2022}.
The large number of clusters that appear in the clustering diagrams in \autoref{fig:clustering_teegarden} serves as a good example of the procedure for evaluating the cluster analysis.
As described in \autoref{subsubsec:cluster_search}, the signals from one year and its integer fractions can interfere with the search for the stellar rotation period (see the top two panels of the figure). Therefore, we focussed on the bottom panel of the figure, where we removed data points from the dataset that were consistent within their uncertainties (i.e. peak width in the GLS) with integer fractions of one year up to 1/6 before applying the clustering algorithm.

The distribution of clusters in the bottom clustering diagram can be explained by three different signals. First, the cluster with the most data points and overall lowest FAP has a period of \SI{\sim 98}{\day} and is consistent with the rotation periods 
in the literature. There is an alias of this signal at \SI{\sim 82}{\day} due to an approximately yearly sampling frequency in the window function of the data. 
Another cluster at \SI{\sim 54}{\day} can likely be attributed to the second harmonic of the presumed rotation period. The period of \SI{29.4}{\day} coincides with the lunar cycle, but is also very close to the third harmonic of the presumed rotation period (see also the analysis of spurious signals in \autoref{subsec:non_activity}). The second-biggest cluster has a period of \SI{\sim 114}{\day} with an alias at \SI{\sim 160}{\day} caused by the yearly sampling. This signal could be related to the spurious RV signal at \SI{\sim175}{\day} that was already discussed by \cite{Zechmeister2019}. Alternatively, it could also be related to the rotation period, since long periods tend to have large uncertainties \citep{Shan2024}. Lastly, there is a third cluster of signals coinciding with the length of the data baseline that corresponds to a long-term trend in the GLS periodograms, and is probably an imprint of a stellar magnetic cycle. Very similar periods were 
independently found by \citet{Fuhrmeister2024} based only on H$\alpha$ and classical periodogram analysis.

For completeness, we note here that some indicators (the RV being the most significant one) show peaks close to periods of the three planets orbiting the star (orbital periods of $\sim$4.9, 11.4, and 26.1 days, for planets b, c, and d, respectively). However, contrary to the other signals discussed above, no significant cluster is observed at any of these periods because these periodicities do not appear in other indicators.

\subsubsection{Application to the whole sample}

\begin{figure}[]
    \centering
    \includegraphics[width=0.4\textwidth]{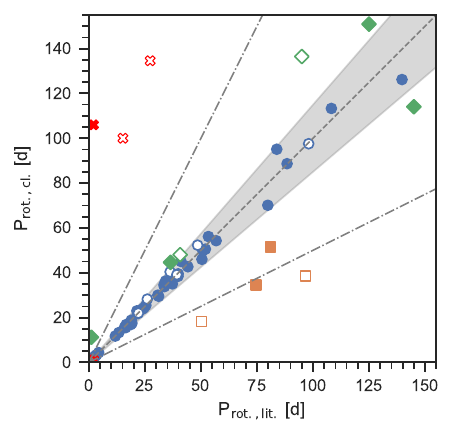}
    \caption{Comparison between rotation periods determined from the clustering algorithm and literature values. Depicted are the results for the signals found with the pre-whitening method in the periodogram analysis. Rotation periods where we found a direct match are depicted by blue circles. Cases where the period from the clustering algorithm is a harmonic or an alias of the literature period are marked by orange squares and green diamonds, respectively. Red crosses highlight stars where the period from the clustering algorithm has no apparent relation to the literature value. Filled symbols depict confirmed rotation periods, while open symbols depict tentative ones, following \citet{Shan2024}. The grey shaded area and dashed line signal a 1-to-1 correlation with a margin of \SI{15}{\percent} of the value. A harmonic correlation (either double or half of the period) is indicated by the grey dashed-dotted lines.}
    \label{fig:comparison_clustering_lit_period}
\end{figure}

We performed an automated search for the stellar rotation periods on the whole sample and compared the results for the naive and pre-whitening methods from the signal search. We removed the signals close to one year and its integer fractions (up to 1/6; i.e. data points consistent with those within their uncertainties). To determine the stellar rotation period, we defined the most likely period, $P_\text{rot,cl}$, to correspond to the cluster with the most data points in it that has a period below \SI{200}{\day}. Alternatively, selecting the cluster with the lowest mean FAP as the rotation period did not change our conclusions, nor did taking the cluster with the overall data point with the lowest FAP. The results for the rotation periods determined in this way are listed in \autoref{tab:rotation_periods}, together with other key properties of the stars. Additional clusters present in the period range below \SI{200}{\day}, which were not selected as the rotation period, were usually connected to the respective rotation period by aliasing or harmonic relations. Only a few cases showed additional clusters at unrelated periods, which could indicate other forms of stellar variability or spurious signals. However, a detailed analysis of these clusters is beyond the scope of this paper.

\cite{Shan2024} provided a review on the rotation periods of 253 stars from the CARMENES survey, most of which measured from photometric data. For our selected stars, they listed 97 stars with rotation periods, $P_\text{rot,lit}$, that are either `secured' or `tentative' (but likely), which we compared with the results from our period search as presented in \autoref{fig:comparison_clustering_lit_period}. We assumed that it was a match if the difference between the period obtained by the clustering algorithm and the literature value was less than \SI{15}{\percent} of the literature period. We recovered 36 (\SI{38}{\percent}) of the rotation periods using the naive approach and 42 (\SI{45}{\percent}) using the pre-whitening for the signal search. 
Taking into account the instances in which the clustering algorithm detected a harmonic or a first-order alias of the literature period (daily or yearly sampling considered), we found additional matches for the periods in 18 cases (\SI{57}{\percent} in total) using the naive approach and in 10 cases (\SI{55}{\percent} in total) using the pre-whitening approach.

The naive method thus appears to detect more rotation-related signals, while the pre-whitening method yields the most direct matches. Furthermore, the two methods differ not only in the number but also in the actual signals that are found. In the naive method, 9 periodicities were detected in stars for which no activity signal was found with the pre-whitening method and, conversely, the pre-whitening method found 7 signals that were not identified with the naive method. Combining both methods, 56 out of 97 stars with known rotation periods were found by the cluster analysis to have activity signals either related to the literature value directly, harmonically, or by aliasing. This combination shows that a careful manual analysis of all clusters that occur, as shown in the previous section for Teegarden's Star, is beneficial.

The biggest difference between the two methods is the number of false-positive detections, namely, the case where a cluster is found, but its period does not match the literature period according to our criteria. The nominal number of false-positives is 19 for the naive method and only 4 for the pre-whitening method. The reason behind this difference is that without pre-whitening, there are often clusters of high order aliases of the period of one year that are not caught by our filtering for the spurious periods (see \autoref{subsec:non_activity} for a more in-depth investigation of those signals). Given the higher number of direct matches and the lower number of false positives, we therefore proceeded with the results from the pre-whitened signal search in the further analysis. 

The majority of the unrecovered literature periods are cases where no clusters of activity indicators were found at all. The false positives occur for stars where the actual period is not present as a cluster, as, for example, in LP~560--035 and/or where the actual rotation period is shorter than \SI{1.5}{\day} -- and is therefore outside the range of our signal search; in such cases, spurious signals at longer periods were determined instead. We note however that for vB~8, with a literature period shorter than \SI{1.5}{\day}, some cluster match a daily alias of the period.

For two additional false positives (G~264--012 and GJ~810\,B), the rotation period is `tentative' according to \cite{Shan2024}. Both have long rotation periods ($P$ \SI{\geq100}{\day}),  which makes the detection inherently more difficult (see the next section and the discussion in \autoref{sec:discussion}). Lastly, HH~And displays a cluster of long-term signals ($P>\SI{200}{\day}$; see also \autoref{subsec:other_activity}), which could lead to erroneous signals.

We also examined the three stars for which a cluster was detected, but no rotation period is known in the literature 
(BD+09~2636, BD-07~3856, and HD~199305; 
highlighted in bold font in \autoref{tab:rotation_periods}). After a visual inspection of the cluster diagrams for each star (see \autoref{app:new_periods}), we deemed all of them to be credible detections and therefore publish them as new estimates of the stellar rotation period for those stars. The periods range from \SI{21}{\day} to \SI{160}{\day}, and the stars cover the spectral types M0.5\,V to M1.0\,V. The period of \SI{160}{\day} determined for HD~199305 seems rather long for such an early M1.0\,V star.
However, the reliability is rather high having four indicators (\He{}, \CaIRTc{}, \CaIRTb{}, \dLW{}) exhibiting FAP lower than~\SI{1}{\percent}.
We further checked the phase-folded data of the relevant activity indicators for these three targets. All display a significant periodicity at the reported periods, which reinforces our confidence on the period values.

\subsubsection{Relations to the stellar properties}
\begin{figure}[t]
    \centering
    \includegraphics[width=0.4\textwidth]{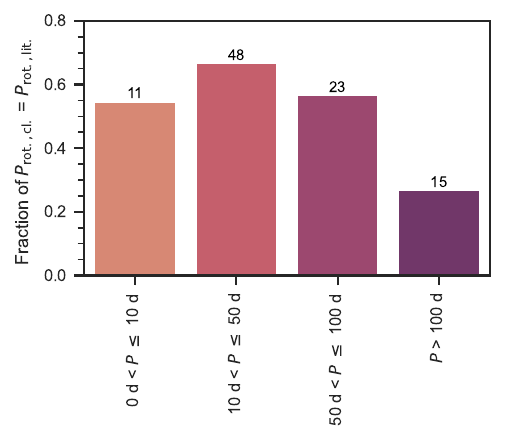}
    \caption{Fraction of stars for which the literature rotation period was recovered as a function of the rotation period. The number on top of each bar denotes the number of stars with a literature rotation period for each category.}
    \label{fig:period_matches_per_period_range}
\end{figure}

\begin{figure*}[t]
    \centering
    \includegraphics[width=0.4\textwidth]{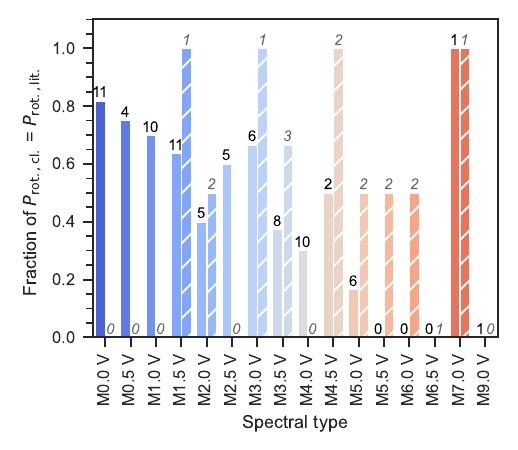}
    \includegraphics[width=0.4\textwidth]{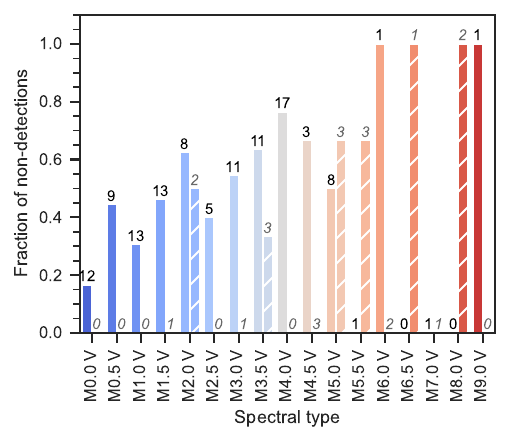}
    \caption{Recovery rate and fraction of non-detections of stellar rotation period as a function of stellar spectral type. \textit{Left}: Fraction of stars for which the literature periods were recovered (either directly or in a harmonic or alias relation). \textit{Right}: Fraction of stars that show no significant clusters (including stars without a literature rotation period). In the plots we distinguish between inactive and active (pEW(\Halpha{}) $<\SI{-0.5}{\angstrom}$) stars by using solid and hatched bars, respectively. In the left panel, the number on top of each bar denotes the number of stars with a known rotation period for each category. In the right panel, it yields the overall number of stars in the selected sample. 
    For better readability, we chose normal black font for the inactive sample and italic grey font for the active sample.}
    \label{fig:comparison_spt}
\end{figure*}

We searched for relations between the signals detected by the clustering algorithm and the stellar properties of our sample. In \autoref{fig:period_matches_per_period_range}, we show the recovery rate of the literature period as a function of the stellar rotation period. We had significantly fewer matches for longer rotation periods, as can be expected since stars with longer rotation periods tend to be less active, implying a minor variability amplitude in the indicators \citep[e.g.][]{Reiners2012b,Newton2017}.

The recovery rate as a function of spectral type is difficult to assess because there are few very late-type M dwarfs in the sample (see left panel in \autoref{fig:comparison_spt}). However, when this is taken together with the number of stars for which the clustering algorithm did not yield any clusters (see right panel in \autoref{fig:comparison_spt}), a consistent picture emerges. For both \Halpha{}-active and inactive stars, the recovery rate and non-detections display a high success rate for early to mid-M dwarfs, but a noticeable drop for stars of spectral types M3.5\,V to M4.0\,V. For later spectral types, the results are very mixed due to the small sample size, with both high recovery and non-detection rates. One reason is that five of the stars with spectral types later than M5.5\,V have known rotation periods shorter than 1.5\,d, which is outside the range of our signal search as described above.

There are other stellar parameters that can be related to our detection of rotation periods.
Recently, \citet{Cortes-Contreras2024} studied the kinematics and its connection to rotation period and velocity, X-ray, near-UV, and H$\alpha$ emission, and magnetic field strength of the whole CARMENES input catalogue, including all our M dwarfs \citep{Alonso-Floriano2015, Reiners2022}.
From the comparison of the list of stars in their study and for which we searched for a $P_{\rm rot,cl}$, we concluded that most of our 136 stars are neither very young nor very old.
The younger stars in our sample rotate generally faster than our 1.5\,d limit, while the oldest ones tend to have very low amplitudes of spectral activity. 
There are, however, a few remarkable exceptions to the intermediate ages, such as two relatively old stars in the Galactic thick disk (V1352~Ori, LP~734--42) and a dozen young star candidates.
Among the last ones, there are bona fide members in young stellar kinematic groups, such as YZ~CMi in $\beta$~Pictoris \citep{Alonso-Floriano2015b} and LP~731--76 in TW~Hya \citep{Gagne2015}, and M dwarfs in the galactic young disk, not ascribed to any group, but with a plethora of youth indicators, such as 1RXS~J114728.8+664405, TYC~ 3529--1437--1, and EV~Lac, just to cite a few \citep{Shkolnik2009,Cortes-Contreras2024}.
We were able to identify between 7 and 16 clusters of activity indicators exactly at the periods reported in the literature for four of the five young stars above.
The fifth young star, namely LP~731--76, has neither a reported period nor a period identified by us, despite its higher X-ray, near-UV, and H$\alpha$ emission and stronger magnetic field than field stars of the same spectral type.
The lack of any rotation period reported may be due to a homogeneous distribution of heterogeneities on the surface of the star.

\subsubsection{Predictive power of the activity indicators}

\begin{figure*}[]
    \centering
    \includegraphics[width=0.9\textwidth]{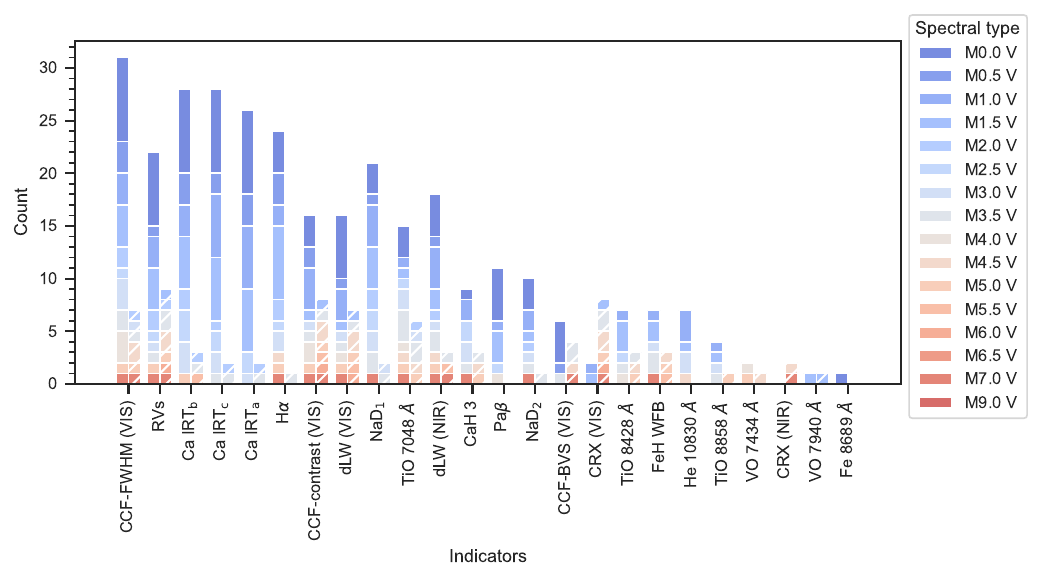}
    \caption{Histogram showing how often  individual indicators were part of  clusters that matched  stellar rotation periods from the literature (either directly or in a harmonic or alias relation). The different colours show the total counts in each bar, differentiated by stellar spectral type. We distinguish between inactive and active (pEW(\Halpha{}) $<\SI{-0.5}{\angstrom}$) stars by using solid and hatched bars respectively.}
    \label{fig:period_matches_per_indicator}
\end{figure*}

Our cluster analysis can inform on the ability of each indicator to trace rotation.  To do so, we investigated the types of activity indicators that formed the clusters corresponding to the stellar rotation signal.
For a general overview, we show in \autoref{fig:period_matches_per_indicator} how frequently each indicator belongs
to the cluster that matches the literature period either directly or is in an alias or harmonic relation, respectively. The plot gives a good indication of the predictive power of the individual indicators regarding the stellar rotation period.
There is however a caveat.
Results may differ if indicators are examined independently of each other \citep[e.g.][]{Schofer2019,Schofer2022,Lafarga2021}. The reason is that in our case the existence of a cluster is a requirement to be counted. This means that an indicator may individually show a signal at the rotation period but not appear in the graph because there are no other indicators showing the same signal.

The \FWHM{} from the VIS channel of CARMENES overall performs best and occurs in 35 of the 52 detected clusters whose period matches the literature period in some form. RV is, as expected among the best performing indicators, especially sensitive for active stars. Similar overall performance is shown by \CaIRT{} and the \Halpha{} indices, which however mostly appear for \Halpha{}-inactive stars. Further, while the \FWHM{}, RVs, and \Halpha{} perform well for all spectral types, a signal of the stellar rotation period in \CaIRT{} is only detected for M dwarfs of early spectral types. The latter holds also for the \NaDdoublet{}, where, interestingly, the \NaDone{} line is performing considerably better than the \NaDtwo{} and has twice the number of detections.

The diagnostic parameters of the \dLW{} of both VIS and NIR channels, VIS-\CONTRAST{}, VIS-\CRX{}, and \BISECTOR{} are independent of the spectral type just like the \FWHM{}. 
However, they have a lower predictive power regarding the stellar rotation. Interestingly, the \dLW{} from both channels perform similarly well, and in general better than the \CRX{} of both channels.

Of the photospheric band indices, \TiOsevenzero{} is particularly remarkable. Excluding diagnostic indicators, \TiOsevenzero{} is the most sensitive for both active and non-active stars. Additionally, in contrast to most of the other indicators, it seems to be overall also sensitive for mid-M dwarfs, where our recovery rate in general was pretty low. The \CRX{} and \Fe{} indicators both trace rotation better for the active than the inactive stars. It is noticeable that the indicator \HeDthree{} was never found in the rotation period clusters and, therefore, does not appear in \autoref{fig:period_matches_per_indicator}. The possible reason for this is that the \HeDthree{} line only shows variability in very active stars, of which there are only a few in our sample.

An interesting application of the clustering algorithm is that it also allows for  an investigation of the relations between the indicators, which we present in \autoref{app:relations_between_ind}. The main result is that indicators of the same kind are likely to appear together in clusters, since they are often affected by activity in the same way and, therefore, show the same signals. This partly explains, for example, why the CCF indicators and \CaIRT{} are particularly frequent in the clusters, as they can form clusters by themselves. It also highlights a strength of the clustering analysis. A good indication of its genuine nature is when a cluster contains several data points from indicators of the same type, which can be easily traced. However, we note than one cannot completely discard similar indicators to be similarly affected by the same biases.

\subsection{Detecting long-term cycles}\label{subsec:other_activity}

\begin{figure}[t]
    \centering
    \includegraphics[width=0.4\textwidth]{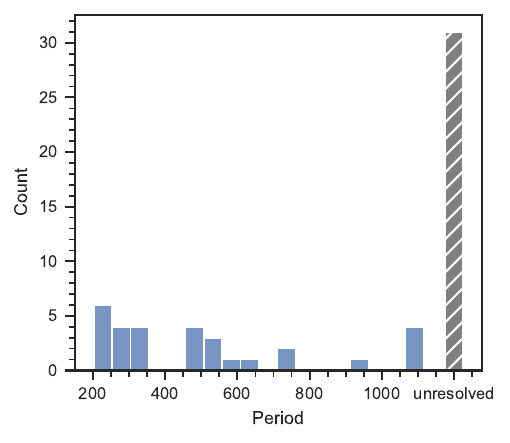}
    \caption{Period distribution of the long-term signals ($P>\SI{200}{\day}$) detected by the automated clustering algorithm.}
    \label{fig:lts_period_distribution}
\end{figure}

\begin{figure}[t]
    \centering
    \includegraphics[width=0.4\textwidth]{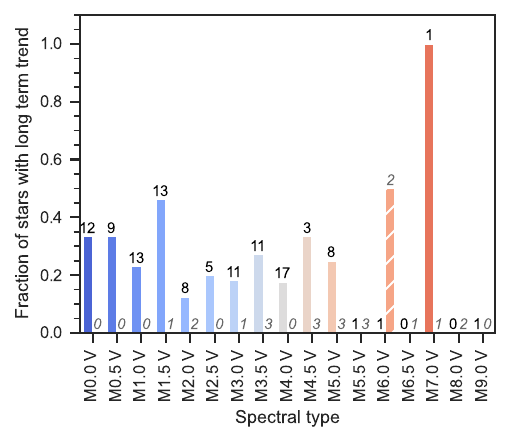}
    \caption{Same as \autoref{fig:comparison_spt}, but for fraction of stars that show long-term signals in the clustering algorithm. 
    }
    \label{fig:lts_fraction_of_detections}
\end{figure}

\begin{figure*}[t]
    \centering
    \includegraphics[width=0.9\textwidth]{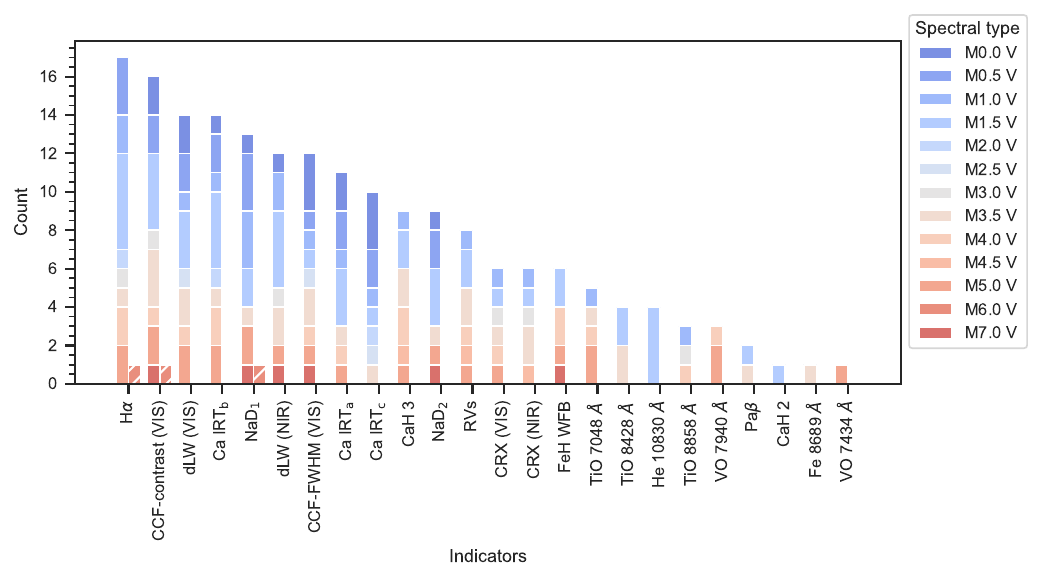}
    \caption{Histogram of the activity indicators that appear in clusters of long-term signals. The different colours show the total counts in each bar, differentiated by stellar spectral type. We distinguish between inactive and active (pEW(\Halpha{}) $<\SI{-0.5}{\angstrom}$) stars by using solid and hatched bars, respectively.}
    \label{fig:lts_per_indicator}
\end{figure*}

In addition to determining the stellar rotation period, we used the automatic clustering algorithm of our sample to look for long-term activity cycles in the data. We considered periodicities above \SI{200}{\day} to be cycle candidates. We found 40 stars (\SI{29}{\percent}) that have clusters with periods greater than \SI{200}{\day} that are unrelated to the periods that we determined as the stellar rotation periods (see \autoref{tab:long_term_signals} for an overview). The majority of these identifications (31) are signals with periods corresponding to the timespan of the observations, which indicates unresolved long-term trends that appear as a plateau for longer periods in the periodograms (see \autoref{fig:lts_period_distribution}). The rest seem to be divided into roughly three groups with periods around \SI{300}{\day}, \SIrange{450}{750}{\day}, and \SIrange{950}{1150}{\day}.

The strong concentration around the multiples of a year is suspicious and suggests that some signals might be false positives. Since a detailed investigation of all our detections as, for example, \cite{Fuhrmeister2023}, would be beyond the scope of this work, we therefore concentrated our further analysis on the unexplained signals. They usually have longer baselines and are more consistent with the measurements of activity cycles from photometry and the $R'_\text{HK}$ index \citep{SuarezMascareno2016,DiezAlonso2019,Fuhrmeister2023}.

As in \autoref{subsec:rotation_periods}, we looked for a possible correlation between detected clusters and stellar spectral type (see \autoref{fig:lts_fraction_of_detections}). Because the overall number of detections is low, any interpretation has to be taken with caution, but in general a lower fraction of the mid-type M dwarfs seems to show long-term trends. The majority of the detections come from inactive stars. There are only two active stars with signals, both late type stars. In \autoref{fig:lts_per_indicator} we show the breakdown of the indicators that are present in the clusters of the long-term trends. Overall, the distribution exhibits some differences in comparison to \autoref{fig:period_matches_per_indicator}. For example, there seems to be no bias of the indicators regarding spectral types. Further, in contrast to the stellar rotation periods, the \FWHM{} is much less frequent in clusters of long-term signals. Also, the \CONTRAST{} and \NaDone{} perform much better for the longer periods than for the shorter rotation periods.


\subsection{Non activity-related signals}
\label{subsec:non_activity}

\subsubsection{Clustering at the lunar cycle}
The alias pair of \SI{27.4}{\day} and \SI{29.4}{\day} occurring in the clustering diagram of Teegarden's star led us to look for similar signals in our entire sample and investigate the possible contamination of our spectra by lunar stray light. We found five stars with clusters appearing at \SIrange{27}{30}{\day}. However, all of these could either be explained as likely harmonics of a known stellar rotation period as in the case of Teegarden's star or were associated with inactive stars that did not show any activity related signals in our analysis and have unconfirmed rotation periods according to \cite{Shan2024}. If, indeed, these inactive stars do not show variability-related signals in the periodogram, this would mean that these signals at \SIrange{27}{30}{\day} could actually be caused by the sampling of the data. This is because RVs are often taken on bright nights, which imprints the lunar cycle into the window function of the sampling. We therefore assumed that contamination by the lunar spectrum has no significant influence on our activity indicators.

\subsubsection{Clustering at the harmonics of one year}
\begin{figure}[t]
    \centering
    \includegraphics[width=0.4\textwidth]{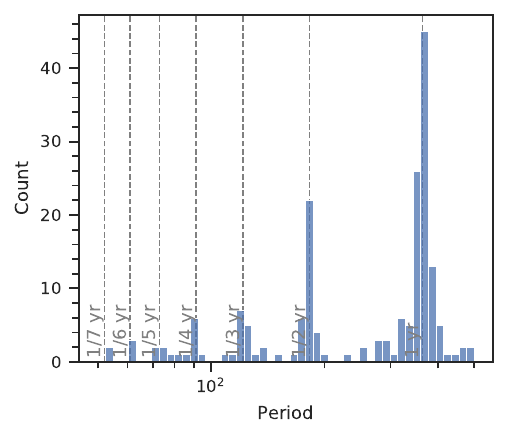}
    \caption{Period distribution of signals occurring around one year and its integer fractions.}
    \label{fig:spu_histogram_periods}
\end{figure}

Signals around the period of one year and its integer fractions occur persistently in the clustering analysis, even when using the pre-whitening method for the signal search. It is present for 120 of the 136 targets in the sample and covers periods ranging from a full year to a sixth or a seventh of a year (see \autoref{fig:spu_histogram_periods}). Though those signals also often appear in periodograms of activity indicators in the literature \citep[e.g.][]{Kossakowski2023,Standing2023, Sartori2023, Lee2023}, they are seldom specifically addressed. 
They are probably not independent harmonic signals, as is often assumed, and could be concluded from the fact that they occur despite our pre-whitening of the periodograms. Instead, we propose that they are aliases of a spurious signal of about one year affecting the data.

The coincidence seen among the view of  harmonics and aliases is explained by the fact that besides the usually dominant sampling frequency of one day in the window function of the data, there is also often a yearly sampling frequency apparent (since most stars are only observable seasonally), which can produce the observed periods\footnote{Aliasing occurs at $f_a=f_u\pm m f_s$ (where $f_a$ is the alias frequency, $f_u$ the underlying signal, $f_s$ the sampling frequency and $m$ is an integer value). For an underlying signal with a period of one year and a yearly sampling, it follows for positive $m$: $f_a = (m+1)/365$, which corresponds to the harmonics of one year in period space}. In principle, the distinguishing feature is that in the case of aliasing, the occurring signals in the clustering analysis depend on the actual dominant frequencies in the window function of the data, and we do not expect an indicator to show up in multiple clusters of the series if we use the pre-whitening method as we do here. In contrast, in the case of harmonics, we expect a fixed ratio between the occurring signals and an indicator can show up in multiple clusters as they are independent of each other.

The vast majority of the clusters in the sample that are related to one year and its integer fractions (hereafter, spurious clusters) are consistent with both explanations. The reason is that in many cases the sampling frequency is very close to one year, so that aliases and harmonics cannot be distinguished within the uncertainties (i.e. the scatter within the cluster or the resolution of the GLS). In addition, for most of the targets, we did not find more than two spurious clusters, which inhibits a meaningful analysis of the repeated occurrence of indicators. We therefore limited our analysis to the few stars whose sampling deviates significantly from the period of one year or that have a particularly large number of clusters related to one year.

In doing so, we found evidence for the presence of both aliasing and harmonics. For example, HD~147379 \citep{Reiners2018a} has an apparent sampling frequency in the window function of the data of $\sim$1/430\SI{}{\per\day} and shows an alias pair at \SI{372}{\day} and \SI{203}{\day} that is inconsistent with being harmonics. On the contrary, BD+11~2576 ($f_s=1/379\SI{}{\per\day}$) shows a regular harmonic series of \SI{366}{\day}, \SI{179}{\day}, \SI{121}{\day}, \SI{91}{\day} and \SI{74}{\day}, where the indicators \CaHthree{}, \He{} and \TiOeighteight{}{} each appear three times in different clusters.
Most importantly, however, we also found a relationship between the occurrence of multiple spurious clusters and the presence of long-term cycles in the clustering analysis. Our resulting hypothesis is that the same indicators appearing in different clusters may not be symptomatic of harmonics, but the result of a non-ideal model in the pre-whitening in the signal search for the stars with long-term trends.

In the case of aliasing, the cause of the underlying signal is unlikely to be a purely sampling effect due to the frequencies in the window function.
Reasons for this are, for example, that the clusters around a year also occur in combination with very strong signals of stellar activity, contrary to the fact that the sampling periods usually only appear for very homogenous data \citep{Dawson2010}. Furthermore, while we would expect a positive correlation between the occurring cluster periods and the sampling frequencies if they were the same, we actually determined a slightly negative correlation (Pearson correlation coefficient of $r=\num{-0.19}$).


\section{Discussion}
\label{sec:discussion}

\subsection{Clustering as a method to determine stellar rotation periods}
Our comparison of the periods obtained from activity indicator clustering with the rotation periods reported in the literature showed that our method can be an effective tool for the determination of stellar rotation periods. We have also shown that pre-whitening the data prior to analysis leads to fewer false-positive detections. However, because the automatic search often finds aliases or harmonics of the literature rotation period, complementary photometric observations and a detailed manual analysis of the occurring cluster periods, like we presented for Teegarden's star, should be used to confirm the results. 
False negative detections (rotation periods identified by one method but not the other) can occur when the rotation period overlaps with an alias frequency (in the case of the naive method) or loss of samples (in the case of the pre-whitening method). Therefore, comparing the naive and pre-whitening results can help detect these false negatives.

In this work, we focussed on a homogeneous analysis of the spectral activity indicators. It would be interesting to extend the clustering analysis to photometry data. Note that the photometry must have comparable temporal resolution as that of the spectral activity indicators, since the method uses a universal distance parameter for all data points in the clustering. In our work, we did not use photometry because it is only available for a sub-sample of our data set and has a wide scatter in temporal resolution \citep{DiezAlonso2019,Shan2024}.

The current implementation of our method to search for stellar rotation periods has some limitations. Since we restrict the analysis to periods longer than 1.5 days, our method is unsuitable for the analysis of fast rotators. Furthermore, the exclusion of signals around one year and its harmonics means that it would also miss any true signals occurring at these periods. Therefore, for stars exhibiting high activity (i.e. likely fast rotators) or yield negative detections, it is recommended to perform manual inspection and signal searches on an unrestricted frequency range. In the case of fast rotators, dense photometric monitoring remains the most efficient and effective way to determine rotation periods, especially in the era of TESS.

Furthermore, the performance of our method is sensitive to the stability, longevity, and intensity contrast of active regions.  Since active regions can be complex and have finite lifetimes, observations over long timespans might trace different active regions. This complexity and time-evolution can result in low significance and non-coherent periodicities in the activity indicators, which in turn can strongly affect the detection of a clear periodicity in the periodograms, especially over long timescales. There is no easy solution to this problem, but one possible mitigation strategy could be to work with selected sub-sets of data exhibiting a roughly constant activity level, if the quantity and time coverage of data permits doing so with minimal loss to signal significance \citep[see e.g.][]{Schofer2022}.

\subsection{Influence of the spectral type and the global activity level}

Using the information from the clustering analysis, we analysed the relations between the different activity indicators and the stellar properties. 
We were able to reproduce a number of results from  previous studies of the activity indicators.

We showed that the number of stars with clusters related to the photometric stellar rotation period is highest for early-type M stars, regardless of the \Halpha{} activity level. There is a notable lower sensitivity of the chromospheric indicators with respect to active stars compared to the diagnostic indicators, which is in agreement with the previous work of \cite{Lafarga2021} and \cite{Schofer2022}. On the contrary, the \dLW{} and \CONTRAST{} perform especially well for \Halpha{}-active stars.

We found a lower detection rate of rotation periods for stars with rotation periods longer than \SI{100}{\day} \citep[e.g.][]{Lafarga2021}. Besides the fact that this relation of activity and rotation periods is well known in the literature \citep[e.g.][]{Astudillo-Defru2017a,Newton2017}, we note that also our methodology has an impact here. For the same amplitude of the signals, the phase coverage is always worse for longer rotation periods than for shorter periods given a data set with the same sampling and timespan (see e.g. \citealt{Jeffers2022} for the impact of different sampling on the periodicities detected for the mid-M dwarf EV\,Lac). This dependency directly affects the power of the signals in the periodograms following the $\hat{\chi}_0^2$ in the denominator of Eq.~59 in \cite{VanderPlas2018}. For a quantitative analysis, a correction for the completeness of the detected signals, as it is for example usually done for retrieving occurrence rates of RV detected stars \citep[e.g.][]{Sabotta2021}, would therefore be necessary.

{\subsection{Dearth of periodicities of mid-M spectral types}

In this work we found a previously unreported decrease in the appearance of periodic signals related to the stellar rotation period for stars with mid-M spectral types, corresponding to the transition from partially to fully convective stars between M3.5~V to M4.5~V. The reconstruction of the large-scale magnetic field geometry using the technique of Zeeman-Doppler imaging shows that mid-M dwarfs typically have magnetic fields that are stronger and more stable than the weaker and more complex fields of early-M dwarfs, while stars with late M spectral types can have either strong and stable or weak and complex fields ~\citep{Morin2008,Morin2010}. The detailed investigation of the stellar activity of the cornerstone mid-M dwarf EV\,Lac  by \cite{Jeffers2022}, using low-resolution Doppler imaging techniques applied to CARMENES spectra over a time span of a few years, is onsistent with these results
is.  We showed that EV\,Lac has a dominant large and stable activity feature, but that there is also evidence of very rapid evolution of small stellar activity features visible in the comparison of several sectors of data from the Transiting Exoplanet Survey Satellite (TESS).

In support of the presence of a component of rapid evolution are the results from  \cite{Yang2023A&A...669A..15Y}, who analysed TESS data spanning just over two years. They determined the flaring rates of more than 13\,000 stars with spectral types spanning late-F to late-M.  They reported that mid-M dwarfs have the highest flaring rates of their stellar sample. A possible explanation for the dearth of periodicities for mid-M dwarfs is the presence of these small-scale activity features that are below the resolution of Zeeman-Doppler imaging but are sufficiently irregular to impact the periodicities of the individual activity lines.
The changes in the geometry of the surface magnetic fields underlying the flares, caused by the transition between dynamo modes, could also be the cause \citep[e.g.][]{Reiners2009,Morin2008,Morin2010,Kitchatinov2014,Yadav2015}.

Another point to consider is the sampling rate of the observations.  As an extension of the investigation of EV~Lac, \cite{Jeffers2022} reported that small sub-sets of data that span not more than a few stellar rotation periods showed a remarkably high detection rate of the rotation period in the stellar activity indices. The method that we have developed in this work also has a great potential as a diagnostic of the complexity of the stellar activity features, and particularly for data sets where the rotation period of the star is densely sampled.  The rapid evolution of the stellar activity indices could indicate that there is a high temporal evolution of the small scale activity features, and in the future we will investigate this further.
}

\subsection{The predictive power of the indicators to trace rotation}

How often each indicator is found in activity-related clusters also allowed us to make statements about their predictive power. The overall best performance regarding the stellar rotation period are the \FWHM{} and the \CaIRT{}. It is also notable that indicators of the same type, such as the three lines of the \CaIRT{}, often show the same signals, which favours the detection of a cluster. We found that the \CaIRT{} is only sensitive for inactive stars with earlier spectral types, as previously reported by \cite{Lafarga2021}. The \TiOsevenzero{} indicator performs best among the photospheric indicators, both for \Halpha{}-active and -inactive stars, and is particularly sensitive to the intermediate-type M dwarfs. This is in agreement with the results of \cite{Schofer2022}, who showed that \TiOsevenzero{} can exhibit a rotational signal over the entire range from \Halpha{}-inactive and slowly rotating to very active and rapidly rotating stars. The fact that they are particularly sensitive to the mid-M dwarfs, in contrast to the chromospheric indicators, can be explained by a combination of two effects. First, the indicators are less sensitive to earlier spectral types because the TiO bands are generally weaker at higher temperatures. Secondly, as shown, the number of activity signals generally decreases for later spectral types, so that the optimum is reached for middle spectral types.

The equal predictive power of the VIS and NIR \dLW{} regarding the stellar rotation period might seem contradictory since the amplitude of activity induced signals in the RV in general decreases with increasing wavelength because of the lower contrast between the spots and the stellar surface \cite[e.g.][]{Desort2007,Reiners2010,Anglada-Escude2013}. By design, the \dLW{} is however specifically sensitive to line-broadening due to the Zeemann effect, which can be particularly strong at NIR wavelength \citep[e.g.][]{Donati2009,Reiners2013,Reiners2022}. The NIR \dLW{} indicator could therefore be a particularly good tracer for changes in the stellar magnetic activity.

Interestingly, the two \NaDdoublet{} lines show a huge difference in their predictive power, the \NaDone{} line being much more efficient than \NaDtwo{} for stars earlier than M4.0\,V. The two resonant lines show a complicated behaviour, since they consist of a strong photospheric absorption component, which gets increasingly shallower for later spectral types, and a chromospheric emission core, which is not present in the least active stars. Moreover, both lines may be affected by strong airglow \citep{Osterbrock1992,Slanger2003}. While the absorption troughs show a similar intensity, which reflects in the indices exhibiting about the same values, the weak emission cores and airglow are about a factor of two stronger in the \NaDtwo{} line, following the $\log{gf}$ ratio obtained from the VALD database \citep{VALD}.
This effect leads to the situation that in many stars an emission core can be identified in the \NaDtwo{} line but not in the \NaDone{} line. We therefore conclude that photospheric variations seem to cause the variation in the \NaDone{} line, while the interplay of photospheric and chromospheric variations leads to a veiling of the periodic signatures in the \NaDtwo{} line.
This complicated behaviour deserves further attention, which is beyond the scope of this study. In previous CARMENES studies, the sum of both lines was typically analysed, which blurred out this effect. A treatment of an average of both lines indices is also usually used in the sparse literature on activity seen in the \NaDdoublet{} \citep[e.g.][]{GomesdaSilva2011}. From our findings, a separate treatment of the two lines is advisable at least for early M dwarfs, making use of the higher sensitivity of the \NaDone{} line in these stars.

Activity indicators appearing in the clusters of long-term signals showed some differences to that of the stellar rotation periods. In addition to the generally smaller sample size, the sensitivity of the activity indicators to the long-term stability of the CARMENES instrument plays a role here. For instance, it is reasonable that the \CONTRAST{} performs best because it remains more stable over long periods of time than the \FWHM{}, which is much more sensitive to changes in the instrument's line spread function caused by, for example, seasonal temperature changes during the year. Further, long-term signals activity indicators based on well-defined spectral lines such as \Halpha{} have an advantage over band indices, such as the TiO indicators, which cover multiple lines that may not behave in an entirely identical manner and thus introduce intrinsic scatter.

\subsection{Clustering at spurious periods}

While contamination from the lunar spectrum does not have a significant impact on our activity indicators, the signals around one year and its integer fractions are ubiquitous. Our hypothesis is that these signals are not likely to be independent harmonic signals, but instead aliases of only one underlying signal affecting the activity indicators. While there is some scatter in the actual period due to the uncertainties in the periodograms, the root of the signal seems to be closely related to a period of one year and does not depend on the sampling of the data. We observed that a pure sinusoid is not sufficient to model the variations in all cases, suggesting that the variability in the activity indicators could be caused by a process that is only quasi-periodic. Possible explanations include variations of the line spread function caused by recurring disturbances on the instrument, such as small annual temperature or pressure changes in the instrument room \citep{Bauer2020}. External influences such as changes in the ambient temperature and, therefore, the chemistry of the atmosphere can also lead to a varying contamination of the spectra. For example, we saw similar signals with a period of about one year for the RVs of CARMENES survey stars, which are caused by contamination by telluric lines \citep[][]{Nagel2023A&A...680A..73N}. In the case of an atmospherical origin, we would however expect these signals to have some dependency on the stellar spectral type that we did not observe. Further, we did not see a clear correlation between the signals in the telluric contaminated RVs and the activity indicators.

Based on these assumptions, pre-whitening of the periodograms is an important and appropriate tool to uncover the actual activity signals in the data. However, because the changes imprinted onto the spectra seem to be only of a quasi-periodic nature, a simple sinusoidal model can still lead to residual signals at the aliases of one year, which needs to be considered during the analysis of the clustering results. Our approach of removing all signals close to one year and its integer fractions from the data before applying the clustering algorithm is a simple way to reduce those residuals.

\subsection{Occurrence of long-term cycles}

The accumulation of the long-term activity cycles around multiples of one year that we found should be a warning sign regarding their veracity. A plausible explanation is that some of the signals are related to spurious signals. The clustering analysis as a method for detecting long-period signals should therefore not be used on its own, but only in combination with other complementary measurements.

The unexplained long-term cycles are preferentially detected in inactive stars. Presumably, this is an observational bias, since active stars are  noisier, which complicates the detection. Apart from this, as described in the previous section for the longest rotation periods, our method of searching for signals with the GLS periodogram is biased towards shorter periods.
As a result, a correction for the completeness of the detected signals would be mandatory in quantitative analysis.

\section{Conclusion}
\label{sec:conclusion}

We present a novel approach to detecting stellar rotation periods and other stellar activity signals in the periodograms of the time series of stellar activity spectral indicators by using a machine learning clustering algorithm. This approach shows promise in identifying spectroscopic modulations related to stellar rotation and can be used to complement rotation period measurements from stellar photometric variability. Analysing the identified clusters makes it possible to combine the information content in many activity indicators simultaneously and also to include signals with higher FAPs in the analysis. It is a way to organize and interpret the multitude of signals occurring in various spectroscopic indicators. Using this method, we demonstrate the successful recovery of 56 rotation periods consistent with those in the literature, and report three new periods.

The benefit of combining several indicators is illustrated by our analysis of the relationships of the indicators to each other, as well as to the stellar properties. The individual indicators have different sensitivities to different aspects of stellar activity. 
A large survey with extensive spectroscopic and temporal coverage, such as CARMENES, provides an excellent opportunity to explore the imprints of stellar activity phenomena on a variety of spectroscopic indicators. The cluster analysis presented here is a powerful tool for investigating relationships between the indicators and stellar activity. This approach will be potentially useful for uncovering other activity-sensitive lines, as well as laying the groundwork for understanding their physical origins.

\begin{acknowledgements}
    The authors thank the anonymous referee for their helpful comments that improved the quality of the manuscript.

    This publication was based on observations collected under the CARMENES Legacy+ project.

    CARMENES is an instrument at the Centro Astron\'omico Hispano en Andaluc\'ia (CAHA) at Calar Alto (Almer\'{\i}a, Spain), operated jointly by the Junta de Andaluc\'ia and the Instituto de Astrof\'isica de Andaluc\'ia (CSIC). The authors wish to express their sincere thanks to all members of the Calar Alto staff for their expert support of the instrument and telescope operation.

    CARMENES was funded by the Max-Planck-Gesellschaft (MPG),
    the Consejo Superior de Investigaciones Cient\'{\i}ficas (CSIC),
    the Ministerio de Econom\'ia y Competitividad (MINECO) and the European Regional Development Fund (ERDF) through projects FICTS-2011-02, ICTS-2017-07-CAHA-4, and CAHA16-CE-3978,
    and the members of the CARMENES Consortium
    (Max-Planck-Institut f\"ur Astronomie,
    Instituto de Astrof\'{\i}sica de Andaluc\'{\i}a,
    Landessternwarte K\"onigstuhl,
    Institut de Ci\`encies de l'Espai,
    Institut f\"ur Astrophysik G\"ottingen,
    Universidad Complutense de Madrid,
    Th\"uringer Landessternwarte Tautenburg,
    Instituto de Astrof\'{\i}sica de Canarias,
    Hamburger Sternwarte,
    Centro de Astrobiolog\'{\i}a and
    Centro Astron\'omico Hispano-Alem\'an),
    with additional contributions by the MINECO,
    the Deutsche Forschungsgemeinschaft through the Major Research Instrumentation Programme and Research Unit FOR2544 ``Blue Planets around Red Stars'',
    the Klaus Tschira Stiftung,
    the states of Baden-W\"urttemberg and Niedersachsen,
    and by the Junta de Andaluc\'{\i}a.

    We acknowledge financial support from a UKRI Future Leader Fellowship grant number MR/S035214/1, the UKRI grant EP/X027562/1, the ERDF “A way of making Europe” through project PID2019-109522GB-C5[1,2,3,4], the DFG priority program SPP 1992 "Exploring the Diversity of Extrasolar Planets (JE 701/5-1), and the Research Council of Norway through the Centres of Excellence funding scheme, project number 332523 (PHAB).

    The analysis of this work has made use of a wide variety of public available software packages that are not referenced in the manuscript: \texttt{scipy} \citep{Virtanen2020}, \texttt{NumPy} \citep{Harris2020}, \texttt{matplotlib} \citep{Hunter2007}, \texttt{tqdm} \citep{daCosta-Luis2019}, \texttt{pandas} \citep{Thepandasdevelopmentteam2020,McKinney2010}, \texttt{seaborn} \citep{Waskom2021}.
    
\end{acknowledgements}

\bibliographystyle{./aa-package/bibtex/aa}
\bibliography{clustering.bib}

\begin{appendix}

\section{Activity indicators used in this work}\label{app:ind_description}

\begin{sidewaystable*}
\centering
\caption{List of activity indicators used in the analysis.}
\label{tab:overview_activity_indicators}
\tabcolsep=0.08cm
{\scriptsize
\begin{tabularx}{\textwidth}{>{\raggedright\arraybackslash}>{\hsize=.15\hsize}X >{\raggedright\arraybackslash}>{\hsize=.08\hsize}X >{\raggedright\arraybackslash}>{\hsize=.12\hsize}X >{\raggedright\arraybackslash}>{\hsize=.28\hsize}X >{\raggedright\arraybackslash}>{\hsize=.28\hsize}X >{\hsize=.3\hsize}X >{\hsize=0.55\hsize}X}
\hline\hline
\multicolumn{7}{c}{\textit{\textbf{Chromospheric line indicators}}\tablefootmark{(d)}:} \\
Indicator         & Central $\lambda$ [\AA] & Central $\Delta\lambda$\tablefootmark{(a)} & Left $\Delta\lambda$\tablefootmark{(b)} & Right $\Delta\lambda$\tablefootmark{(b)} & Examined in\tablefootmark{(c)} & Description \\
\hline
\noalign{\smallskip}
\Halpha{}   & 6562.8  & $\pm40$~km\,s$^{-1}$ & $-500$ to $-300$~km\,s$^{-1}$        & $300$ to $500$~km\,s$^{-1}$          & \underline{Ze18}, T-O18, Fu19, Sch19, La21, Je22, Sch22, Fu23 & \multirow[t]{9}{=}{Chromospheric lines can be characterised by their emission flux in the line core. This emission flux can be measured from the line's pseudo-equivalent width (pEW) or line index, as defined in Sch19 and Ze18, respectively, for the data used here. The pEW is the equivalent width of the line with respect to a pseudo-continuum, {with} the spectrum of a similar non-active star {subtracted from it} to remove photospheric contributions. The line index is the ratio of mean flux around the line centre to the flux in reference bandpasses on either sides of the line, regions assumed to not be affected by stellar activity. Both pEW and line index measure the 0th moment of the line (i.e. depth, area, or integrated flux). We present the different lines of the Ca\,{\sc ii} infrared triplet and the \NaDdoublet{} in different rows.} \\
\CaIRTa{}   & 8498.0  & $\pm15$~km\,s$^{-1}$ & $\pm40$~km\,s$^{-1}$ from 8492.0~\AA & $\pm40$~km\,s$^{-1}$ from 8504.0~\AA & \textquotedbl                                                 & \\
\CaIRTb{}   & 8542.1  & $\pm15$~km\,s$^{-1}$ & $-300$ to $-200$~km\,s$^{-1}$        & 250 to 350~km\,s$^{-1}$              & \textquotedbl                                                 & \\
\CaIRTc{}   & 8662.1  & $\pm15$~km\,s$^{-1}$ & $-300$ to $-200$~km\,s$^{-1}$        & 200 to 300~km\,s$^{-1}$              & \textquotedbl                                                 & \\
\NaDone{}   & 5895.9  & $\pm15$~km\,s$^{-1}$ & $\pm40$~km\,s$^{-1}$ from 5885.0~\AA & $\pm40$~km\,s$^{-1}$ from 5892.9~\AA & \underline{Ze18}, Fu19, Sch19, Je22, Sch22                    & \\
\NaDtwo{}   & 5889.9  & $\pm15$~km\,s$^{-1}$ & $\pm40$~km\,s$^{-1}$ from 5892.9~\AA & $\pm40$~km\,s$^{-1}$ from 5905.0~\AA & \textquotedbl                                                 & \\
\HeDthree{} & 5875.6  & $\pm$2.5~\AA & $\pm$1.25~\AA~from 5871.1~\AA   & $\pm$1.25~\AA~from 5881.1~\AA  & \underline{Sch19}, Fu20, Je22, Sch22                          & \\
\He{}       & 10830.3 & $\pm$0.25~\AA & $\pm$1.00~\AA~from 10818.0~\AA & $\pm$1.00~\AA~from 10871.7~\AA & \textquotedbl                                                 & \\
\Pabeta{}   & 12818.1 & $\pm$0.25~\AA & $\pm$1.00~\AA~from 12813.5~\AA & $\pm$1.00~\AA~from 12822.5~\AA & \underline{Sch19}, Je22, Sch22                                & \\
\noalign{\medskip} 
\multicolumn{7}{c}{\textit{\textbf{Photospheric absorption band indices}}:} \\
Indicator        & \multicolumn{2}{c}{Numerator $\Delta\lambda$ [\AA]} & \multicolumn{2}{c}{Denominator $\Delta\lambda$ [\AA]} & Examined in\tablefootmark{(c)}                             & Description \\
\hline
\CaHtwo{}        & \multicolumn{2}{c}{$6815.1-6817.1$} & \multicolumn{2}{c}{$6811.1-6813.1$} & \underline{Sch21}, Je22              & \multirow[t]{8}{=}{Photospheric bands can cover multiple spectral orders and are blended with other spectral features, which challenges the measurement of their pEW or line index, as typically done with chromospheric lines. Instead, photospheric band indices can be defined as the ratio of mean fluxes in two different small ranges on either side of the band head (Sch19). Similarly to chromospheric line indicators, these indices measure the 0th moment of the band absorption lines.} \\
\CaHthree{}      & \multicolumn{2}{c}{$6972.1-6975.1$} & \multicolumn{2}{c}{$7045.1-7048.1$} & \textquotedbl                        & \\
\TiOsevenzero{}  & \multicolumn{2}{c}{$7054.1-7058.1$} & \multicolumn{2}{c}{$7044.1-7048.1$} & \underline{Sch19}, Sch22, Je22, Fu23 & \\
\TiOeightfour{}  & \multicolumn{2}{c}{$8433.7-8437.7$} & \multicolumn{2}{c}{$8428.2-8432.2$} & \textquotedbl                        & \\
\TiOeighteight{} & \multicolumn{2}{c}{$8860.1-8862.1$} & \multicolumn{2}{c}{$8856.6-8858.6$} & \textquotedbl                        & \\
\VOsevenfour{}   & \multicolumn{2}{c}{$7433.9-7434.9$} & \multicolumn{2}{c}{$7432.3-7433.3$} & \underline{Sch19}, Sch22, Je22       & \\
\VOsevenine{}    & \multicolumn{2}{c}{$7939.5-7941.5$} & \multicolumn{2}{c}{$7933.8-7935.8$} & \textquotedbl                        & \\
\WFB{}           & \multicolumn{2}{c}{$9895.3-9904.3$} & \multicolumn{2}{c}{$9884.3-9893.3$} & \underline{Sch21}, Je22              & \\
\noalign{\medskip}
\end{tabularx}
\begin{tabularx}{\textwidth}{>{\raggedright\arraybackslash}>{\hsize=0.15\hsize}X >{\raggedright\arraybackslash}>{\hsize=0.165\hsize}X >{\raggedright\arraybackslash}>{\hsize=0.685\hsize}X}
\multicolumn{3}{c}{\textit{\textbf{Diagnostic indicators}}:} \\
{Indicator}       & {Examined in\tablefootmark{(c)}}                             & {Description} \\
\hline
{chromatic index (\CRX) \tablefootmark{(e)}}                & {\underline{Ze18}, T-O18, La21, Je22, Sch22} & {The \CRX{} measures wavelength-dependent changes in the RVs. It is defined as the slope between the order-wise RVs and the average logarithmic wavelength of the order (Ze18).} \\
{differential line width (\dLW)\tablefootmark{(e)}}         & {\underline{Ze18}, T-O18, La21, Je22}        & {The \dLW{} measures differential variations in the absorption line widths of the observed spectrum with respect to a spectral template when performing a template matching minimisation (Ze18). It traces similar variations as the CCF FWHM and measures the 2nd moment of the line profile (i.e. variance).}  \\
{CCF Full width at half maximum (\FWHM)\tablefootmark{(f)}} & {\underline{La20}, La21, Je22, Sch22}        & {\multirow[t]{3}{=}{The CCF represents an average shape of the stellar absorption lines, hence, its profile can be used to trace activity effects. The FWHM and the contrast are typically derived from a Gaussian fit to the CCF. The FWHM measures changes in the average width of the absorption lines (2nd moment of the CCF). The contrast traces the 0th moment of the CCF. The BIS is the difference between the average RV of the top region of the CCF profile and the average RV of the bottom region{, originally defined by \citet{Queloz2001}}. It traces asymmetries in the line profile, so it measures the 3rd moment of the profile (skewness).}} \\
{CCF bisector inverse slope (\BISECTOR)\tablefootmark{(f)}} & {\textquotedbl}                              &  \\
{\CONTRAST\tablefootmark{(f)}}                              & {\textquotedbl}                              &  \\
\hline
\end{tabularx}
}
\tablebib{\scriptsize Ze18: \cite{Zechmeister2018}; Sch19: \cite{Schofer2019}; Sch21: \cite{Schofer2021}; La20: \cite{Lafarga2020}; T-O18: \cite{Tal-Or2018}; Fu19: \cite{Fuhrmeister2019a}; La21: \cite{Lafarga2021}; Je22: \cite{Jeffers2022}; Sch22: \cite{Schofer2022}; Fu23: \cite{Fuhrmeister2023}.}
\tablefoot{
\scriptsize
\tablefoottext{a}{{Passband $\Delta\lambda$ used to integrate the flux coming from the central line, indicated in km\,s$^{-1}$ from the central line, unless otherwise specified (i.e. indicated in \AA~for the indicators computed following Sch19).}}
\tablefoottext{b}{{Left and right bandpasses $\Delta\lambda$ at either side of the central line used as continuum reference, indicated in km\,s$^{-1}$ from the central line, unless otherwise specified (i.e. indicated in km\,s$^{-1}$ from a different reference wavelength for the \NaDdoublet~lines, and indicated in \AA~from a different reference wavelength for the indicators computed following Sch19).}}
\tablefoottext{c}{Listing only works by the CARMENES consortium, where underline indicates the reference where the indicator is defined. {Quote (\textquotedbl) indicates that the references are the same as in the row above.}}  
\tablefoottext{d}{The indicators calculated following \cite{Zechmeister2018} are dimensionless indices, while the indicators from Sch19, Sch22 are pEWs (in \si{\angstrom}) calculated after subtracting an inactive reference star spectrum.}
\tablefoottext{e}{Calculated for the VIS and NIR channel.}
\tablefoottext{f}{Calculated only for the VIS channel.}
}
\end{sidewaystable*}

\FloatBarrier

\section{Relations between the activity indicators}\label{app:relations_between_ind}

\begin{figure*}[!ht]
    \centering
    \includegraphics[width=0.9\textwidth]{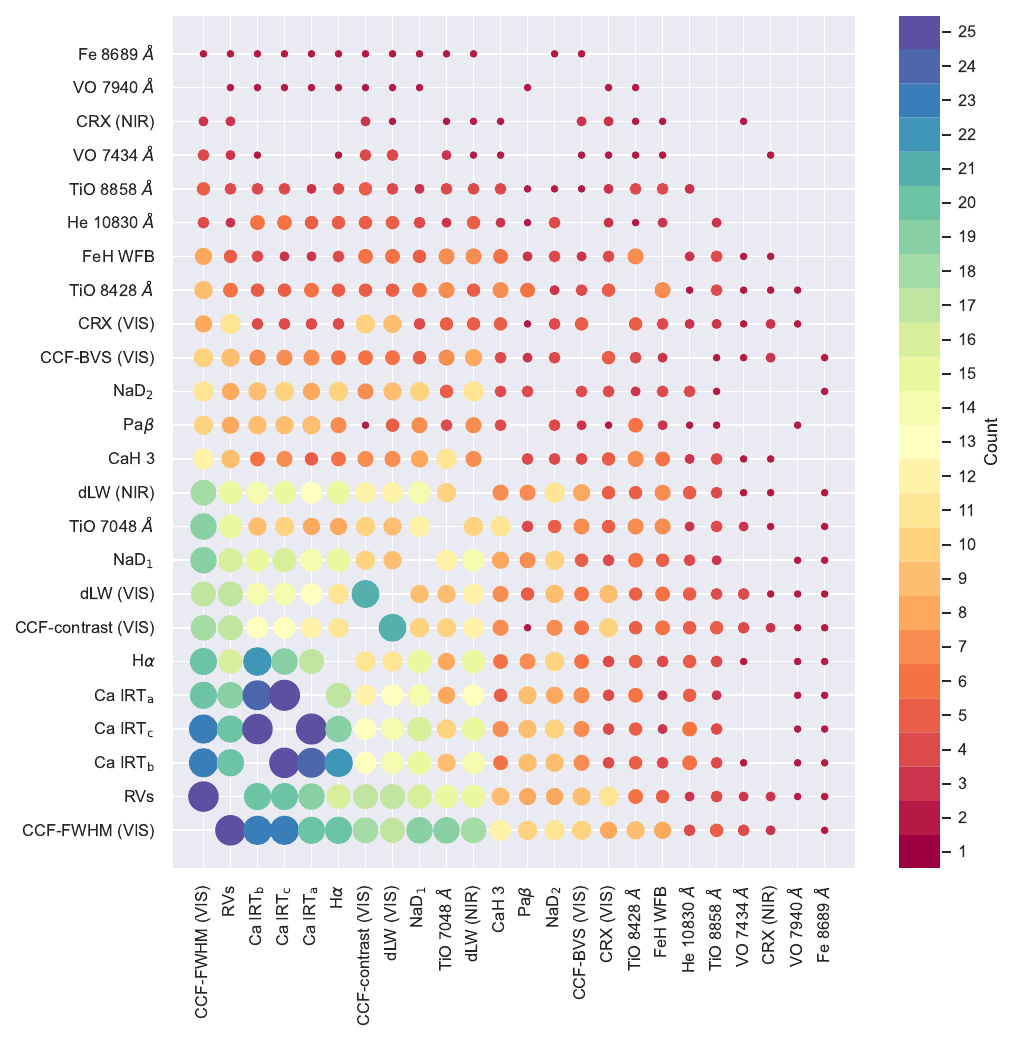}\\
    \caption{Relations between the occurrence of the indicators. The dots represent how often the individual indicators appear together in a cluster that matches the stellar rotation period (either directly or in a harmonic or alias relation). The size and colour of the dots depict the actual incidence: The bigger or bluer a dot, the more often the two indicators appear together.}
    \label{fig:relations_between_ind}
\end{figure*}

An advantage of the clustering method is that it additionally allowed us to directly examine the correlations between the occurrence of the activity indicators. In \autoref{fig:relations_between_ind} we show individually the incidence with which each indicator is present together with the other indicators in the clusters matching the stellar rotation period. Mainly, the diagram reflects the general frequency of the indicators from \autoref{fig:period_matches_per_indicator}. It can be observed that indicators of the same type also frequently occur together, as can be expected. This is especially true for the \CaIRT{}, as well as the diagnostic indicators (CCF parameters, \dLW{} and \CRX{}), which presumably are so abundant because they intrinsically show the same signals and can thus form their own clusters without other indicators occurring. A correlated occurrence can be further observed for the diagnostic and chromospheric indicators in general. Regarding individual indicators, it is worth mentioning that for example the TiO band indicators as photospheric indicators are found to show the highest correlation with other heavy molecules such as \CaHthree{} and the \WFB{}, which is likely due to their higher abundance in stars with lower temperatures (both, physically and in the frequency in the clusters). Noteworthy are the \dLW{}, \NaDone{}, \TiOsevenzero{}, which show interesting relations to other indicators.

Contrary to what one might expect, the \dLW{} shows the highest correlation with the \CONTRAST{} and not with the \FWHM{} (second most abundant). Both \FWHM{} and \dLW{} trace the variance of the line profiles, which means they represent the second moment of the line profile. The \CONTRAST{} measures the depth of the profile, and therefore represents the zeroth moment. The \dLW, however, also has a zeroth moment component \citep{Zechmeister2018,Jeffers2022}, while the \CONTRAST{} and \FWHM{} are expected to be correlated because the equivalent width of the CCF is conserved to zeroth order, which could explain the observed correlations. Furthermore, the \TiOeighteight{} index interestingly appears more often together with the \CRX{} than other line indicators. Not unexpectedly, the \NaDone{} index occurs particularly frequently with other indicators with a strong chromospheric component, such as the \CaIRT{}. Yet, it also shows a particularly strong correlation with the NIR-\dLW{}. Both \CaHthree{} and \WFB{} present a notable relation to the occurrence of the \TiOsevenzero{} band indicator, with the \WFB{} in particular also correlating with \CaHthree{}, and generally occurring more frequently with the three TiO band indices.

    {\onecolumn
        \section{Results tables}

        \begin{landscape}
            \scriptsize
            \begin{longtable}{llSSSSlllSp{3.8cm}}
                \caption{Stellar rotation periods determined from the clustering algorithm for the whole sample.} \label{tab:rotation_periods}                                                                                                                                                                                                                                                                                                                                                                                                                                                                                            \\
                \hline\hline
                {Karmn}               & {Name}                & {$P_\text{rot.,cl.}$ [d]} & {Std. [d]\tablefootmark{(a)}}     & {GLS uncert. [d]\tablefootmark{(a)}} & {$P_\text{rot.,lit.}$ [d]} & {Ref.$_\text{lit.}$} & {Reliability\tablefootmark{(b)}} & {Match\tablefootmark{(c)}} & {\#Samples} & {Indicators\tablefootmark{(d)}}                                                                                                                                                                                                                                                                                                                                        \\
                \hline
                \endfirsthead
                \caption[]{Continued.}                                                                                                                                                                                                                                                                                                                                                                                                                                                                                                         \\
                \hline\hline
                {Karmn}               & {Name}                & {$P_\text{rot.,cl.}$ [d]} & {Std. [d]\tablefootmark{(a)}}     & {GLS uncert. [d]\tablefootmark{(a)}} & {$P_\text{rot.,lit.}$ [d]} & {Ref.$_\text{lit.}$} & {Reliability\tablefootmark{(b)}} & {Match\tablefootmark{(c)}} & {\#Samples} & {Indicators\tablefootmark{(d)}}                                                                                                                                                                                                                                                                                                                                        \\
                \hline
                \endhead
                \hline
                \multicolumn{11}{l}{\tablebib{DA19: \cite{DiezAlonso2019}; SM23: \cite{SuarezMascareno2023}; SM18: \cite{SuarezMascareno2023}; Sh24: \cite{Shan2024}; New18: \cite{Newton2018};  Laf21: \cite{Lafarga2021}; New16: \cite{Newton2016}; Luq22: \cite{Luque2022}; Sto20b: \cite{Stock2020b}; SM17: \cite{SuarezMascareno2017}; Mor10: \cite{Morin2010}; Diaz19: \cite{Diaz2019}; Ste16: \cite{Stelzer2016}; Luq18: \cite{Luque2018}; SM15: \cite{SuarezMascareno2015}; TP19: \cite{Toledo-Padron2019}; Ama21: \cite{Amado2021}; Wat06: \cite{Watson2006}, Don23: \cite{Donati2023}}}                                                           \\
                \multicolumn{11}{l}{\tablefoot{Rotation periods determined in this work are highlighted in bold font.                                                                                                                                                                                                                                                                                                                                                                                                                                                                                                                     \\ \tablefoottext{a}{Standard deviations and GLS uncertainties smaller then \SI{0.00}{\day} where conservatively set to \SI{0.01}{\day}} \tablefoottext{b}{according to \cite{Shan2024}} \tablefoottext{c}{YA: yearly alias match; DA: daily alias match; H: harmonic match; D: direct match} \tablefoottext{d}{sorted by minimum FAP}}}. \\
                \endfoot
                \hline
                \multicolumn{11}{l}{\tablebib{DA19: \cite{DiezAlonso2019}; SM23: \cite{SuarezMascareno2023}; SM18: \cite{SuarezMascareno2023}; Sh24: \cite{Shan2024}; New18: \cite{Newton2018};  Laf21: \cite{Lafarga2021}; New16: \cite{Newton2016}; Luq22: \cite{Luque2022}; Sto20b: \cite{Stock2020b}; SM17: \cite{SuarezMascareno2017}; Mor10: \cite{Morin2010}; Diaz19: \cite{Diaz2019}; Ste16: \cite{Stelzer2016}; Luq18: \cite{Luque2018}; SM15: \cite{SuarezMascareno2015}; TP19: \cite{Toledo-Padron2019}; Ama21: \cite{Amado2021}; Wat06: \cite{Watson2006}, Don23: \cite{Donati2023}}}                                                           \\
                \multicolumn{11}{l}{\tablefoot{Rotation periods determined in this work are highlighted in bold font.                                                                                                                                                                                                                                                                                                                                                                                                                                                                                                                     \\ \tablefoottext{a}{Standard deviations and GLS uncertainties smaller then \SI{0.00}{\day} where conservatively set to \SI{0.01}{\day}} \tablefoottext{b}{according to \cite{Shan2024}} \tablefoottext{c}{YA: yearly alias match; DA: daily alias match; H: harmonic match; D: direct match} \tablefoottext{d}{sorted by minimum FAP}}}. \\
                \endlastfoot
                J00051+457 & BD+44 4548 & {\dots} & {\dots} & {\dots} & 15.37 & DA19 & dubious & {\dots} & {\dots} & {\dots} \\
J00067-075 & GJ 1002 & 161.28 & 2.93 & 0.75 & 126.00 & SM23 & dubious & YA & 6 & CCF-FWHM (VIS), dLW (NIR), TiO 7048 \AA, H$\alpha$, Ca IRT$_\mathrm{c}$, TiO 8428 \AA \\
J00183+440 & GX And & 41.13 & 0.30 & 0.01 & 45.00 & SM18 & secure & D, YA & 4 & RVs, Ca IRT$_\mathrm{c}$, CCF-FWHM (VIS), Ca IRT$_\mathrm{a}$ \\
J00184+440 & GQ And & 108.20 & 1.72 & 0.16 & 113.30 & Don23 & secure & D & 7 & TiO 7048 \AA, CCF-FWHM (VIS), NaD$_1$, dLW (NIR), FeH WFB, NaD$_2$, Ca IRT$_\mathrm{c}$ \\
J00389+306 & Wolf 1056 & 18.31 & 0.05 & 0.01 & 50.20 & DA19 & tentative & H & 3 & Ca IRT$_\mathrm{a}$, Ca IRT$_\mathrm{b}$, NaD$_1$ \\
J01025+716 & Ross 318 & 51.98 & 1.56 & 0.07 & 50.52 & Sh24 & secure & D, YA & 10 & H$\alpha$, TiO 7048 \AA, CCF-FWHM (VIS), dLW (VIS), CCF-contrast (VIS), Ca IRT$_\mathrm{b}$, Ca IRT$_\mathrm{c}$, CaH$_3$, TiO 8858 \AA, He 10830 \AA \\
J01026+623 & BD+61 195 & 18.91 & 0.11 & 0.01 & 18.82 & Sh24 & secure & D, YA & 8 & H$\alpha$, Ca IRT$_\mathrm{b}$, TiO 7048 \AA, Ca IRT$_\mathrm{a}$, Ca IRT$_\mathrm{c}$, CCF-FWHM (VIS), NaD$_1$, RVs \\
J01048-181 & GJ 1028 & {\dots} & {\dots} & {\dots} & 143.20 & New18 & tentative & {\dots} & {\dots} & {\dots} \\
J01125-169 & YZ Cet & 79.82 & 2.09 & 0.11 & 70.10 & Sh24 & secure & D, YA & 6 & CCF-FWHM (VIS), dLW (VIS), RVs, CRX (VIS), CCF-FWHM (VIS), CCF-contrast (VIS) \\
J02002+130 & TZ Ari & 1.96 & 0.01 & 0.01 & 2.00 & Sh24 & secure & D, YA, DA & 4 & RVs, CCF-BVS (VIS), CRX (VIS), TiO 7048 \AA \\
J02015+637 & G 244-047 & {\dots} & {\dots} & {\dots} & {\dots} & {\dots} & {\dots} & {\dots} & {\dots} & {\dots} \\
J02123+035 & BD+02 348 & {\dots} & {\dots} & {\dots} & 17.11 & Sh24 & dubious & {\dots} & {\dots} & {\dots} \\
J02222+478 & BD+47 612 & 31.13 & 0.08 & 0.01 & 29.50 & DA19 & secure & D, YA & 4 & Ca IRT$_\mathrm{b}$, H$\alpha$, Ca IRT$_\mathrm{a}$, CCF-FWHM (VIS) \\
J02362+068 & BX Cet & {\dots} & {\dots} & {\dots} & {\dots} & {\dots} & {\dots} & {\dots} & {\dots} & {\dots} \\
J02442+255 & VX Ari & {\dots} & {\dots} & {\dots} & 38.70 & DA19 & tentative & {\dots} & {\dots} & {\dots} \\
J02530+168 & Teegarden's Star & 98.05 & 1.30 & 0.16 & 97.56 & Laf21 & tentative & D & 12 & CCF-contrast (VIS), CCF-FWHM (VIS), dLW (VIS), dLW (NIR), FeH WFB, NaD$_2$, CaH$_3$, TiO 7048 \AA, NaD$_1$, RVs, CCF-contrast (VIS), H$\alpha$ \\
J03133+047 & CD Cet & 139.71 & 7.09 & 0.54 & 126.20 & New16 & secure & D & 9 & TiO 7048 \AA, dLW (VIS), H$\alpha$, CCF-contrast (VIS), CCF-FWHM (VIS), dLW (NIR), Ca IRT$_\mathrm{b}$, RVs, VO 7434 \AA \\
J03181+382 & HD 275122 & {\dots} & {\dots} & {\dots} & 77.20 & DA19 & tentative & {\dots} & {\dots} & {\dots} \\
J03463+262 & HD 23453 & 10.07 & 0.01 & 0.01 & 16.40 & Sh24 & dubious & H & 4 & H$\alpha$, Ca IRT$_\mathrm{b}$, Ca IRT$_\mathrm{a}$, Ca IRT$_\mathrm{c}$ \\
J04153-076 & omi02 Eri C & 137.41 & 3.97 & 1.34 & 8.56 & Sh24 & dubious & No & 3 & dLW (VIS), TiO 7048 \AA, Pa$\beta$ \\
J04290+219 & BD+21 652 & 25.42 & 0.42 & 0.01 & 25.40 & DA19 & secure & D, YA & 16 & dLW (VIS), Ca IRT$_\mathrm{c}$, Ca IRT$_\mathrm{b}$, H$\alpha$, Ca IRT$_\mathrm{a}$, CCF-contrast (VIS), CCF-FWHM (VIS), Ca IRT$_\mathrm{c}$, dLW (NIR), RVs, Ca IRT$_\mathrm{b}$, NaD$_1$, CCF-FWHM (VIS), dLW (VIS), CCF-BVS (VIS), Ca IRT$_\mathrm{a}$ \\
J04376+528 & BD+52 857 & 16.31 & 0.04 & 0.01 & 15.47 & Laf21 & secure & D, YA & 7 & dLW (VIS), Ca IRT$_\mathrm{c}$, Ca IRT$_\mathrm{a}$, Ca IRT$_\mathrm{b}$, RVs, Pa$\beta$, Ca IRT$_\mathrm{c}$ \\
J04588+498 & BD+49 1280 & 19.21 & 0.05 & 0.01 & 17.46 & Sh24 & secure & D, YA & 7 & Ca IRT$_\mathrm{a}$, Ca IRT$_\mathrm{c}$, Ca IRT$_\mathrm{b}$, RVs, dLW (VIS), CCF-contrast (VIS), CCF-FWHM (VIS) \\
J05033-173 & LP 776-046 & {\dots} & {\dots} & {\dots} & {\dots} & {\dots} & {\dots} & {\dots} & {\dots} & {\dots} \\
J05127+196 & GJ 192 & {\dots} & {\dots} & {\dots} & 32.70 & Sh24 & tentative & {\dots} & {\dots} & {\dots} \\
J05314-036 & HD 36395 & 33.68 & 0.40 & 0.01 & 33.80 & DA19 & secure & D, YA & 8 & Ca IRT$_\mathrm{c}$, Ca IRT$_\mathrm{b}$, Ca IRT$_\mathrm{a}$, H$\alpha$, RVs, CCF-FWHM (VIS), Pa$\beta$, dLW (NIR) \\
J05365+113 & V2689 Ori & 11.79 & 0.02 & 0.01 & 11.75 & Sh24 & secure & D, YA & 14 & CCF-FWHM (VIS), H$\alpha$, RVs, Ca IRT$_\mathrm{a}$, dLW (NIR), Ca IRT$_\mathrm{b}$, Ca IRT$_\mathrm{c}$, NaD$_1$, CCF-BVS (VIS), TiO 7048 \AA, dLW (VIS), CCF-contrast (VIS), NaD$_2$, Fe 8689 \AA \\
J05415+534 & HD 233153 & {\dots} & {\dots} & {\dots} & 17.39 & Sh24 & tentative & {\dots} & {\dots} & {\dots} \\
J05421+124 & V1352 Ori & {\dots} & {\dots} & {\dots} & {\dots} & {\dots} & {\dots} & {\dots} & {\dots} & {\dots} \\
J06011+595 & G 192-013 & 83.82 & 1.05 & 0.08 & 95.1 & Don23 & secure & D & 5 & TiO 7048 \AA, RVs, CCF-contrast (VIS), dLW (VIS), NaD$_1$ \\
J06024+498 & G 192-015 & {\dots} & {\dots} & {\dots} & 105.00 & DA19 & secure & {\dots} & {\dots} & {\dots} \\
J06103+821 & GJ 226 & {\dots} & {\dots} & {\dots} & 44.60 & DA19 & tentative & {\dots} & {\dots} & {\dots} \\
J06105-218 & HD  42581 & {\dots} & {\dots} & {\dots} & 27.30 & DA19 & secure & {\dots} & {\dots} & {\dots} \\
J06371+175 & HD 260655 & {\dots} & {\dots} & {\dots} & 37.50 & Luq22 & secure & {\dots} & {\dots} & {\dots} \\
J06396-210 & LP 780-032 & {\dots} & {\dots} & {\dots} & 79.15 & New18 & tentative & {\dots} & {\dots} & {\dots} \\
J06548+332 & Wolf 294 & 53.54 & 0.17 & 0.01 & 122.00 & Sto20b & dubious & H & 3 & RVs, NaD$_1$, CCF-BVS (VIS) \\
J07274+052 & Luyten's Star & 95.12 & 0.79 & 0.04 & 93.50 & SM17 & dubious & D & 4 & RVs, dLW (VIS), H$\alpha$, dLW (VIS) \\
J07319+362N & BL Lyn & {\dots} & {\dots} & {\dots} & 16.40 & DA19 & tentative & {\dots} & {\dots} & {\dots} \\
J07403-174 & LP 783-002 & {\dots} & {\dots} & {\dots} & {\dots} & {\dots} & {\dots} & {\dots} & {\dots} & {\dots} \\
J07446+035 & YZ CMi & 2.78 & 0.01 & 0.01 & 2.78 & Sh24 & secure & D, YA & 16 & CCF-BVS (VIS), TiO 7048 \AA, RVs, CCF-FWHM (VIS), CRX (VIS), TiO 8428 \AA, CaH$_3$, CRX (NIR), VO 7434 \AA, CCF-contrast (VIS), dLW (VIS), FeH WFB, RVs, CCF-BVS (VIS), CCF-FWHM (VIS), CRX (VIS) \\
J08161+013 & GJ 2066 & 48.06 & 0.41 & 0.01 & 40.70 & DA19 & tentative & YA & 5 & Ca IRT$_\mathrm{b}$, H$\alpha$, NaD$_1$, CCF-FWHM (VIS), RVs \\
J08413+594 & LP 090-018 & 88.41 & 0.70 & 0.04 & 88.65 & Laf21 & secure & D & 5 & dLW (VIS), dLW (NIR), CCF-contrast (VIS), FeH WFB, Ca IRT$_\mathrm{b}$ \\
J08536-034 & LP 666-009 & {\dots} & {\dots} & {\dots} & 0.46 & Sh24 & secure & {\dots} & {\dots} & {\dots} \\
J09144+526 & HD 79211 & 16.67 & 0.03 & 0.01 & 16.88 & Sh24 & secure & D, YA & 10 & Ca IRT$_\mathrm{c}$, Ca IRT$_\mathrm{b}$, Ca IRT$_\mathrm{a}$, CCF-FWHM (VIS), RVs, H$\alpha$, dLW (NIR), NaD$_1$, CCF-BVS (VIS), Pa$\beta$ \\
J09411+132 & Ross 85 & {\dots} & {\dots} & {\dots} & {\dots} & {\dots} & {\dots} & {\dots} & {\dots} & {\dots} \\
J09425+700 & GJ 360 & {\dots} & {\dots} & {\dots} & 21.00 & DA19 & tentative & {\dots} & {\dots} & {\dots} \\
J09428+700 & GJ 362 & 24.57 & 0.03 & 0.01 & 24.33 & Sh24 & secure & D, YA & 3 & CCF-FWHM (VIS), NaD$_1$, TiO 7048 \AA \\
J09561+627 & BD+63 869 & 18.68 & 0.10 & 0.01 & 16.88 & Sh24 & secure & D, YA & 7 & RVs, NaD$_1$, CCF-FWHM (VIS), Pa$\beta$, CaH$_3$, TiO 7048 \AA, RVs \\
J10122-037 & AN Sex & 21.42 & 0.01 & 0.01 & 23.00 & Sh24 & secure & D, YA & 5 & RVs, Ca IRT$_\mathrm{c}$, Ca IRT$_\mathrm{b}$, Ca IRT$_\mathrm{a}$, H$\alpha$ \\
J10289+008 & BD+01 2447 & {\dots} & {\dots} & {\dots} & 31.72 & Sh24 & secure & {\dots} & {\dots} & {\dots} \\
J10482-113 & LP 731-058 & {\dots} & {\dots} & {\dots} & 1.50 & Mor10 & secure & {\dots} & {\dots} & {\dots} \\
J10508+068 & EE Leo & {\dots} & {\dots} & {\dots} & 64.00 & DA19 & tentative & {\dots} & {\dots} & {\dots} \\
J10564+070 & CN Leo & 2.70 & 0.01 & 0.01 & 2.70 & DA19 & secure & D, YA & 4 & RVs, CRX (VIS), CCF-contrast (VIS), dLW (VIS) \\
J10584-107 & LP 731-076 & {\dots} & {\dots} & {\dots} & {\dots} & {\dots} & {\dots} & {\dots} & {\dots} & {\dots} \\
J11000+228 & Ross 104 & {\dots} & {\dots} & {\dots} & 53.17 & Sh24 & secure & {\dots} & {\dots} & {\dots} \\
J11026+219 & DS Leo & {\dots} & {\dots} & {\dots} & 14.26 & Sh24 & secure & {\dots} & {\dots} & {\dots} \\
J11033+359 & Lalande 21185 & 53.33 & 0.19 & 0.01 & 56.15 & Diaz19 & secure & D, YA & 4 & TiO 8428 \AA, H$\alpha$, Ca IRT$_\mathrm{a}$, FeH WFB \\
J11055+435 & WX UMa & {\dots} & {\dots} & {\dots} & 0.78 & Mor10 & secure & {\dots} & {\dots} & {\dots} \\
J11110+304W & HD  97101B & {\dots} & {\dots} & {\dots} & {\dots} & {\dots} & {\dots} & {\dots} & {\dots} & {\dots} \\
J11302+076 & K2-18 & 44.62 & 1.85 & 0.12 & 36.40 & DA19 & secure & YA & 4 & H$\alpha$, Ca IRT$_\mathrm{b}$, CaH$_3$, Ca IRT$_\mathrm{c}$ \\
J11417+427 & Ross 1003 & {\dots} & {\dots} & {\dots} & 71.50 & DA19 & dubious & {\dots} & {\dots} & {\dots} \\
J11474+667 & 1RXS J114728.8+664405 & 13.36 & 0.01 & 0.01 & 13.30 & Sh24 & secure & D, YA & 10 & RVs, TiO 7048 \AA, TiO 8428 \AA, CRX (VIS), CCF-contrast (VIS), CaH$_3$, CCF-FWHM (VIS), TiO 8858 \AA, dLW (VIS), FeH WFB \\
J11476+786 & GJ 445 & {\dots} & {\dots} & {\dots} & {\dots} & {\dots} & {\dots} & {\dots} & {\dots} & {\dots} \\
J11477+008 & FI Vir & {\dots} & {\dots} & {\dots} & 112.80 & Ste16 & secure & {\dots} & {\dots} & {\dots} \\
J11509+483 & GJ 1151 & 150.92 & 3.12 & 0.16 & 125.00 & DA19 & secure & YA & 4 & TiO 7048 \AA, dLW (NIR), H$\alpha$, He 10830 \AA \\
J11511+352 & BD+36 2219 & {\dots} & {\dots} & {\dots} & 22.80 & DA19 & secure & {\dots} & {\dots} & {\dots} \\
J12100-150 & LP 734-032 & {\dots} & {\dots} & {\dots} & 79.30 & Sh24 & tentative & {\dots} & {\dots} & {\dots} \\
J12123+544S & HD 238090 & 38.42 & 0.14 & 0.01 & 96.70 & Sto20b & tentative & H & 6 & dLW (VIS), Ca IRT$_\mathrm{c}$, Ca IRT$_\mathrm{b}$, CCF-contrast (VIS), Ca IRT$_\mathrm{a}$, CCF-FWHM (VIS) \\
J12230+640 & Ross 690 & {\dots} & {\dots} & {\dots} & 32.90 & DA19 & dubious & {\dots} & {\dots} & {\dots} \\
\bfseries J12312+086 & \bfseries BD+09 2636 & \bfseries 21.42 & \bfseries 0.07 & \bfseries 0.01 & \bfseries {\dots} & \bfseries This work & \bfseries {\dots} & \bfseries {\dots} & \bfseries 3 & \bfseries H$\alpha$, Ca IRT$_\mathrm{c}$, Ca IRT$_\mathrm{a}$ \\
J12479+097 & Wolf 437 & {\dots} & {\dots} & {\dots} & {\dots} & {\dots} & {\dots} & {\dots} & {\dots} & {\dots} \\
J13229+244 & Ross 1020 & 136.46 & 2.59 & 0.95 & 95.00 & Luq18 & tentative & YA & 3 & TiO 7048 \AA, CCF-FWHM (VIS), CCF-contrast (VIS) \\
J13299+102 & BD+11 2576 & 30.78 & 0.08 & 0.01 & 30.00 & SM17 & secure & D, YA & 6 & Ca IRT$_\mathrm{b}$, Ca IRT$_\mathrm{c}$, CCF-contrast (VIS), Ca IRT$_\mathrm{a}$, CCF-FWHM (VIS), H$\alpha$ \\
J13457+148 & HD 119850 & 48.51 & 1.86 & 0.04 & 52.30 & SM15 & tentative & D, YA & 10 & Ca IRT$_\mathrm{b}$, H$\alpha$, Ca IRT$_\mathrm{a}$, Pa$\beta$, Ca IRT$_\mathrm{c}$, Ca IRT$_\mathrm{b}$, TiO 8428 \AA, NaD$_1$, VO 7940 \AA, H$\alpha$ \\
J14257+236E & BD+24 2733B & {\dots} & {\dots} & {\dots} & 17.60 & DA19 & dubious & {\dots} & {\dots} & {\dots} \\
J14257+236W & BD+24 2733A & {\dots} & {\dots} & {\dots} & 111.00 & DA19 & tentative & {\dots} & {\dots} & {\dots} \\
\bfseries J14307-086 & \bfseries BD-07 3856 & \bfseries 54.48 & \bfseries 0.50 & \bfseries 0.03 & \bfseries {\dots} & \bfseries This work & \bfseries {\dots} & \bfseries {\dots} & \bfseries 6 & \bfseries Ca IRT$_\mathrm{c}$, dLW (VIS), Ca IRT$_\mathrm{a}$, CCF-contrast (VIS), Pa$\beta$, Ca IRT$_\mathrm{b}$ \\
J14321+081 & LP 560-035 & 2.02 & 0.01 & 0.01 & 1.11 & Sh24 & secure & No & 3 & NaD$_2$, NaD$_1$, TiO 8428 \AA \\
J14342-125 & HN Lib & {\dots} & {\dots} & {\dots} & 94.77 & Sh24 & secure & {\dots} & {\dots} & {\dots} \\
J15194-077 & HO Lib & {\dots} & {\dots} & {\dots} & 132.50 & SM15 & tentative & {\dots} & {\dots} & {\dots} \\
J15218+209 & OT Ser & 3.37 & 0.01 & 0.01 & 3.35 & Sh24 & secure & D, YA & 6 & CCF-contrast (VIS), RVs, CRX (VIS), dLW (VIS), VO 7940 \AA, RVs \\
J16167+672N & EW Dra & 36.24 & 0.09 & 0.01 & 40.40 & Don23 & tentative & D &  5 & NaD$_1$, NaD$_2$, Ca IRT$_\mathrm{b}$, CCF-FWHM (VIS), H$\alpha$ \\
J16167+672S & HD 147379 & 22.06 & 0.07 & 0.01 & 22.00 & Laf21 & tentative & D, YA & 7 & Ca IRT$_\mathrm{a}$, Ca IRT$_\mathrm{b}$, RVs, Ca IRT$_\mathrm{c}$, CCF-BVS (VIS), CCF-FWHM (VIS), TiO 7048 \AA \\
J16303-126 & V2306 Oph & {\dots} & {\dots} & {\dots} & 119.00 & DA19 & secure & {\dots} & {\dots} & {\dots} \\
J16555-083 & vB 8 & 11.19 & 0.02 & 0.01 & 1.09 & Laf21 & secure & DA & 7 & RVs, CCF-contrast (VIS), CCF-BVS (VIS), dLW (NIR), CCF-FWHM (VIS), CRX (NIR), CRX (VIS) \\
J16581+257 & BD+25 3173 & 23.92 & 0.19 & 0.02 & 23.80 & DA19 & secure & D, YA & 4 & Ca IRT$_\mathrm{b}$, Ca IRT$_\mathrm{c}$, Ca IRT$_\mathrm{a}$, He 10830 \AA \\
J17033+514 & G 203-042 & {\dots} & {\dots} & {\dots} & {\dots} & {\dots} & {\dots} & {\dots} & {\dots} & {\dots} \\
J17052-050 & Wolf 636 & {\dots} & {\dots} & {\dots} & 50.20 & DA19 & tentative & {\dots} & {\dots} & {\dots} \\
J17115+384 & Wolf 654 & {\dots} & {\dots} & {\dots} & 62.60 & DA19 & tentative & {\dots} & {\dots} & {\dots} \\
J17303+055 & BD+05 3409 & 33.51 & 0.24 & 0.01 & 34.60 & Sh24 & secure & D, YA & 10 & H$\alpha$, Ca IRT$_\mathrm{a}$, CCF-FWHM (VIS), Ca IRT$_\mathrm{b}$, Ca IRT$_\mathrm{c}$, NaD$_2$, dLW (VIS), dLW (NIR), RVs, Pa$\beta$ \\
J17378+185 & BD+18 3421 & 37.24 & 0.78 & 0.01 & 35.02 & Sh24 & secure & D, YA & 8 & He 10830 \AA, dLW (VIS), Ca IRT$_\mathrm{c}$, CCF-contrast (VIS), dLW (NIR), NaD$_1$, NaD$_2$, Ca IRT$_\mathrm{a}$ \\
J17578+046 & Barnard's Star & 114.06 & 2.30 & 0.06 & 145.00 & TP19 & secure & YA & 9 & CaH$_3$, TiO 7048 \AA, dLW (NIR), FeH WFB, TiO 7048 \AA, CCF-BVS (VIS), TiO 8858 \AA, TiO 8428 \AA, CCF-FWHM (VIS) \\
J18027+375 & GJ 1223 & {\dots} & {\dots} & {\dots} & 123.80 & New16 & tentative & {\dots} & {\dots} & {\dots} \\
J18051-030 & HD 165222 & 34.16 & 0.21 & 0.01 & 127.80 & SM15 & dubious & No & 4 & dLW (NIR), Ca IRT$_\mathrm{c}$, NaD$_1$, Ca IRT$_\mathrm{b}$ \\
J18165+048 & G 140-51 & {\dots} & {\dots} & {\dots} & {\dots} & {\dots} & {\dots} & {\dots} & {\dots} & {\dots} \\
J18174+483 & TYC 3529-1437-1 & 16.00 & 0.07 & 0.01 & 15.83 & Sh24 & secure & D, YA & 7 & Ca IRT$_\mathrm{b}$, Ca IRT$_\mathrm{a}$, RVs, Ca IRT$_\mathrm{c}$, TiO 7048 \AA, RVs, CCF-FWHM (VIS) \\
J18224+620 & GJ 1227 & {\dots} & {\dots} & {\dots} & {\dots} & {\dots} & {\dots} & {\dots} & {\dots} & {\dots} \\
J18346+401 & LP 229-017 & {\dots} & {\dots} & {\dots} & 40.20 & DA19 & tentative & {\dots} & {\dots} & {\dots} \\
J18409-133 & BD-13 5069 & 26.05 & 0.08 & 0.01 & 28.23 & Sh24 & tentative & D, YA & 4 & dLW (VIS), Ca IRT$_\mathrm{c}$, CCF-contrast (VIS), Ca IRT$_\mathrm{a}$ \\
J18482+076 & G 141-036 & {\dots} & {\dots} & {\dots} & 2.76 & DA19 & secure & {\dots} & {\dots} & {\dots} \\
J19072+208 & HD 349726 & {\dots} & {\dots} & {\dots} & 3.80 & DA19 & dubious & {\dots} & {\dots} & {\dots} \\
J19169+051N & V1428 Aql & 50.43 & 1.17 & 0.08 & 46.00 & DA19 & secure & D, YA & 10 & Ca IRT$_\mathrm{b}$, Ca IRT$_\mathrm{a}$, Pa$\beta$, NaD$_1$, Ca IRT$_\mathrm{c}$, TiO 7048 \AA, CaH$_3$, CCF-FWHM (VIS), TiO 8428 \AA, RVs \\
J19169+051S & V1298 Aql & {\dots} & {\dots} & {\dots} & 23.60 & DA19 & dubious & {\dots} & {\dots} & {\dots} \\
J19255+096 & LSPM J1925+0938 & {\dots} & {\dots} & {\dots} & {\dots} & {\dots} & {\dots} & {\dots} & {\dots} & {\dots} \\
J19346+045 & BD+04 4157 & 21.76 & 0.06 & 0.01 & 21.79 & Sh24 & secure & D, YA & 10 & Ca IRT$_\mathrm{c}$, Ca IRT$_\mathrm{b}$, dLW (NIR), Ca IRT$_\mathrm{a}$, dLW (VIS), CCF-FWHM (VIS), H$\alpha$, Pa$\beta$, NaD$_2$, CCF-BVS (VIS) \\
J20260+585 & Wolf 1069 & 156.30 & 8.68 & 0.46 & 57.70 & DA19 & dubious & No & 7 & CCF-BVS (VIS), FeH WFB, TiO 7048 \AA, He 10830 \AA, CaH$_3$, He {\sc i} D$_3$, CCF-FWHM (VIS) \\
J20305+654 & GJ 793 & {\dots} & {\dots} & {\dots} & 32.80 & DA19 & secure & {\dots} & {\dots} & {\dots} \\
J20336+617 & GJ 1254 & {\dots} & {\dots} & {\dots} & 12.60 & DA19 & tentative & {\dots} & {\dots} & {\dots} \\
J20450+444 & BD+44 3567 & 39.23 & 0.10 & 0.01 & 39.12 & Sh24 & tentative & D, YA & 4 & NaD$_1$, Ca IRT$_\mathrm{c}$, CCF-FWHM (VIS), H$\alpha$ \\
J20525-169 & LP 816-060 & {\dots} & {\dots} & {\dots} & 67.60 & DA19 & dubious & {\dots} & {\dots} & {\dots} \\
\bfseries J20533+621 & \bfseries HD 199305 & \bfseries 159.61 & \bfseries 4.53 & \bfseries 0.52 & \bfseries {\dots} & \bfseries This work & \bfseries {\dots} & \bfseries {\dots} & \bfseries 6 & \bfseries He 10830 \AA, Ca IRT$_\mathrm{c}$, Ca IRT$_\mathrm{b}$, dLW (VIS), Ca IRT$_\mathrm{a}$, CCF-contrast (VIS) \\
J20556-140S & GJ 810 B & 27.34 & 0.01 & 0.01 & 134.60 & New18 & tentative & No & 4 & dLW (VIS), NaD$_2$, NaD$_1$, CCF-contrast (VIS) \\
J21164+025 & LSPM J2116+0234 & 44.15 & 0.49 & 0.03 & 42.68 & Sh24 & secure & D, YA & 8 & TiO 7048 \AA, RVs, dLW (NIR), CCF-FWHM (VIS), NaD$_1$, Ca IRT$_\mathrm{a}$, CaH$_3$, Ca IRT$_\mathrm{c}$ \\
J21221+229 & TYC 2187-512-1 & 39.75 & 0.31 & 0.01 & 38.40 & Sh24 & secure & D, YA & 14 & dLW (NIR), Ca IRT$_\mathrm{b}$, Ca IRT$_\mathrm{c}$, Ca IRT$_\mathrm{a}$, CCF-FWHM (VIS), H$\alpha$, NaD$_1$, RVs, NaD$_2$, dLW (VIS), CaH$_3$, He 10830 \AA, CCF-contrast (VIS), CRX (VIS) \\
J21348+515 & Wolf 926 & 56.77 & 0.33 & 0.04 & 54.30 & DA19 & secure & D, YA & 3 & He 10830 \AA, Ca IRT$_\mathrm{b}$, FeH WFB \\
J21463+382 & LSPM J2146+3813 & {\dots} & {\dots} & {\dots} & {\dots} & {\dots} & {\dots} & {\dots} & {\dots} & {\dots} \\
J21466+668 & G 264-012 & 15.17 & 0.04 & 0.01 & 100.00 & Ama21 & tentative & No & 3 & CaH$_3$, dLW (VIS), Ca IRT$_\mathrm{c}$ \\
J22021+014 & BD+00 4810 & {\dots} & {\dots} & {\dots} & 29.50 & DA19 & dubious & {\dots} & {\dots} & {\dots} \\
J22057+656 & G 264-18 A & {\dots} & {\dots} & {\dots} & 120.50 & Laf21 & tentative & {\dots} & {\dots} & {\dots} \\
J22096-046 & BD-05 5715 & {\dots} & {\dots} & {\dots} & 39.20 & SM15 & secure & {\dots} & {\dots} & {\dots} \\
J22115+184 & Ross 271 & 34.23 & 0.16 & 0.01 & 36.30 & DA19 & secure & D, YA & 8 & CCF-FWHM (VIS), Ca IRT$_\mathrm{c}$, NaD$_1$, NaD$_2$, Ca IRT$_\mathrm{b}$, RVs, dLW (NIR), H$\alpha$ \\
J22125+085 & Wolf 1014 & {\dots} & {\dots} & {\dots} & {\dots} & {\dots} & {\dots} & {\dots} & {\dots} & {\dots} \\
J22137-176 & LP 819-052 & {\dots} & {\dots} & {\dots} & 116.40 & Ste16 & tentative & {\dots} & {\dots} & {\dots} \\
J22252+594 & G 232-070 & {\dots} & {\dots} & {\dots} & 64.60 & DA19 & secure & {\dots} & {\dots} & {\dots} \\
J22330+093 & BD+08 4887 & 37.66 & 0.11 & 0.01 & 37.80 & Sh24 & secure & D, YA & 9 & Ca IRT$_\mathrm{c}$, H$\alpha$, RVs, Ca IRT$_\mathrm{a}$, Ca IRT$_\mathrm{b}$, dLW (NIR), NaD$_1$, TiO 8858 \AA, CCF-contrast (VIS) \\
J22468+443 & EV Lac & 4.37 & 0.01 & 0.01 & 4.35 & Sh24 & secure & D, YA & 16 & RVs, dLW (VIS), CCF-FWHM (VIS), TiO 7048 \AA, CCF-contrast (VIS), dLW (NIR), H$\alpha$, CCF-BVS (VIS), CaH$_3$, Ca IRT$_\mathrm{c}$, Ca IRT$_\mathrm{b}$, CRX (VIS), NaD$_1$, Ca IRT$_\mathrm{a}$, TiO 8428 \AA, NaD$_2$ \\
J22503-070 & BD-07 5871 & {\dots} & {\dots} & {\dots} & {\dots} & {\dots} & {\dots} & {\dots} & {\dots} & {\dots} \\
J22532-142 & IL Aqr & 51.39 & 0.42 & 0.02 & 81.00 & DA19 & secure & H & 3 & Pa$\beta$, TiO 8428 \AA, CCF-FWHM (VIS) \\
J22565+165 & HD 216899 & 39.85 & 0.50 & 0.01 & 39.50 & DA19 & secure & D, YA & 25 & dLW (VIS), Ca IRT$_\mathrm{b}$, Ca IRT$_\mathrm{c}$, RVs, Ca IRT$_\mathrm{a}$, dLW (NIR), CCF-contrast (VIS), NaD$_2$, CCF-FWHM (VIS), dLW (VIS), Ca IRT$_\mathrm{b}$, Ca IRT$_\mathrm{c}$, Pa$\beta$, Ca IRT$_\mathrm{a}$, NaD$_1$, H$\alpha$, He 10830 \AA, FeH WFB, Pa$\beta$, TiO 8858 \AA, CCF-FWHM (VIS), CCF-contrast (VIS), dLW (NIR), TiO 8428 \AA, CRX (VIS) \\
J23216+172 & LP 462-027 & 34.46 & 0.38 & 0.01 & 74.70 & DA19 & secure & H & 4 & CCF-contrast (VIS), dLW (VIS), CCF-FWHM (VIS), VO 7434 \AA \\
J23245+578 & BD+57 2735 & {\dots} & {\dots} & {\dots} & 36.48 & Sh24 & secure & {\dots} & {\dots} & {\dots} \\
J23351-023 & GJ 1286 & {\dots} & {\dots} & {\dots} & 88.92 & New18 & dubious & {\dots} & {\dots} & {\dots} \\
J23381-162 & G 273-093 & 47.21 & 0.66 & 0.02 & 61.66 & Wat06 & dubious & YA & 3 & Pa$\beta$, TiO 8858 \AA, CaH$_3$ \\
J23419+441 & HH And & 2.07 & 0.01 & 0.01 & 106.00 & DA19 & secure & No & 3 & NaD$_2$, H$\alpha$, NaD$_1$ \\
J23492+024 & BR Psc & 50.83 & 0.35 & 0.01 & 49.90 & SM18 & secure & D, YA & 8 & dLW (NIR), CCF-FWHM (VIS), TiO 7048 \AA, CaH$_3$, TiO 8428 \AA, Pa$\beta$, NaD$_1$, FeH WFB \\
J23505-095 & LP 763-012 & {\dots} & {\dots} & {\dots} & {\dots} & {\dots} & {\dots} & {\dots} & {\dots} & {\dots} \\
            \end{longtable}
        \end{landscape}


        \begin{landscape}
            \scriptsize
            \begin{longtable}{llSSSSSSSp{5cm}}
                \caption{Long term signals in the data. Signals with periods that are unresolved are reported with the timespan of the data as minimum values.} \label{tab:long_term_signals}                                                                                                                                                                                                                                                                                                                                                                                                                                    \\
                \hline\hline
                {Karmn}     & {Name}           & {$P_\text{lt.,cl.}$ [d]} & {{$P_\text{lt.,Fu23}$ [d]}} & {$P_\text{rot.,cl.}$ [d]} & {$P_\text{rot.,lit.}$ [d]} & {\#Samples} & {timespan [d]} & {RV rms [\si{\meter\per\second}]} & {Indicators\tablefootmark{(a)}}                                                                                                                                                                                                                                                                                                                                                            \\
                \hline
                \endfirsthead
                \caption[]{Continued.}                                                                                                                                                                                                                                                                                                                                                                                                                                                                \\
                \hline\hline
                {Karmn}     & {Name}           & {$P_\text{lt.,cl.}$ [d]} & {{$P_\text{lt.,Fu23}$ [d]}} & {$P_\text{rot.,cl.}$ [d]} & {$P_\text{rot.,lit.}$ [d]} & {\#Samples} & {timespan [d]} & {RV rms [\si{\meter\per\second}]} & {Indicators\tablefootmark{(a)}}                                                                                                                                                                                                                                                                                                                                                            \\
                \hline
                \endhead
                \hline
                \multicolumn{10}{r}{\tablefoot{\tablefoottext{a}{sorted by minimum FAP}}}
                \endfoot
                \hline
                \multicolumn{10}{r}{\tablefoot{\tablefoottext{a}{sorted by minimum FAP}}}
                \endlastfoot
                J00183+440 & GX And & 468 & {\dots} & 41.13 & 45.00 & 3 & 1502.04 & 3.03 & TiO 7048 \AA, Ca IRT$_\mathrm{b}$, CCF-FWHM (VIS) \\
J00183+440 & GX And & >1502 & {\dots} & 41.13 & 45.00 & 3 & 1502.04 & 3.03 & CCF-FWHM (VIS), CRX (VIS), NaD$_1$ \\
J00184+440 & GQ And & >1445 & {\dots} & 108.20 & {\dots} & 13 & 1445.99 & 2.51 & CRX (NIR), NaD$_2$, CCF-FWHM (VIS), CRX (VIS), RVs, NaD$_1$, H$\alpha$, dLW (VIS), TiO 8428 \AA, CCF-contrast (VIS), CaH$_3$, dLW (NIR), Pa$\beta$ \\
J00184+440 & GQ And & 304 & {\dots} & 108.20 & {\dots} & 6 & 1445.99 & 2.51 & RVs, dLW (NIR), CCF-FWHM (VIS), VO 7940 \AA, Fe 8689 \AA, FeH WFB \\
J01026+623 & BD+61 195 & >779 & {\dots} & 18.91 & 18.82 & 7 & 779.99 & 5.53 & CCF-contrast (VIS), dLW (VIS), dLW (NIR), Ca IRT$_\mathrm{b}$, H$\alpha$, Ca IRT$_\mathrm{a}$, CaH$_3$ \\
J02123+035 & BD+02 348 & >1563 & {\dots} & {\dots} & 17.11 & 3 & 1563.67 & 2.03 & He 10830 \AA, Ca IRT$_\mathrm{b}$, CRX (VIS) \\
J02222+478 & BD+47 612 & >718 & {\dots} & 31.13 & 29.50 & 3 & 719.00 & 4.70 & NaD$_2$, H$\alpha$, NaD$_1$ \\
J02530+168 & Teegarden's Star & >1385 & {\dots} & 98.05 & 97.56 & 6 & 1385.35 & 2.98 & CCF-contrast (VIS), CCF-FWHM (VIS), NaD$_1$, FeH WFB, dLW (NIR), NaD$_2$ \\
J04290+219 & BD+21 652 & >749 & 657.00 & 25.42 & 25.40 & 8 & 749.03 & 4.55 & CCF-contrast (VIS), H$\alpha$, NaD$_1$, dLW (VIS), Ca IRT$_\mathrm{c}$, Ca IRT$_\mathrm{b}$, Ca IRT$_\mathrm{a}$, NaD$_2$ \\
J04588+498 & BD+49 1280 & >1464 & {\dots} & 19.21 & 17.46 & 4 & 1464.92 & 8.95 & CCF-contrast (VIS), Ca IRT$_\mathrm{c}$, CCF-FWHM (VIS), NaD$_1$ \\
J06105-218 & HD  42581 & >1421 & {\dots} & {\dots} & 27.30 & 8 & 1421.09 & 5.09 & Ca IRT$_\mathrm{b}$, Ca IRT$_\mathrm{c}$, Ca IRT$_\mathrm{a}$, H$\alpha$, CCF-FWHM (VIS), dLW (VIS), CCF-contrast (VIS), NaD$_1$ \\
J06371+175 & HD 260655 & 741 & {\dots} & {\dots} & 37.50 & 4 & 2310.80 & 3.41 & He 10830 \AA, H$\alpha$, CCF-FWHM (VIS), NaD$_1$ \\
J06371+175 & HD 260655 & >2310 & {\dots} & {\dots} & 37.50 & 3 & 2310.80 & 3.41 & CCF-contrast (VIS), dLW (VIS), NaD$_2$ \\
J06548+332 & Wolf 294 & 646 & {\dots} & 53.54 & 122.00 & 6 & 2498.02 & 3.71 & NaD$_1$, Ca IRT$_\mathrm{b}$, TiO 7048 \AA, FeH WFB, Ca IRT$_\mathrm{a}$, CCF-FWHM (VIS) \\
J06548+332 & Wolf 294 & >2498 & {\dots} & 53.54 & 122.00 & 3 & 2498.02 & 3.71 & H$\alpha$, dLW (NIR), CRX (NIR) \\
J07274+052 & Luyten's Star & 1092 & 1022.00 & 95.12 & 93.50 & 8 & 2498.09 & 3.17 & H$\alpha$, RVs, Ca IRT$_\mathrm{c}$, RVs, CRX (VIS), CCF-contrast (VIS), He 10830 \AA, dLW (VIS) \\
J07274+052 & Luyten's Star & 293 & 1022.00 & 95.12 & 93.50 & 3 & 2498.09 & 3.17 & TiO 8428 \AA, CRX (VIS), NaD$_1$ \\
J07274+052 & Luyten's Star & 551 & 1022.00 & 95.12 & 93.50 & 3 & 2498.09 & 3.17 & RVs, He {\sc i} D$_3$, RVs \\
J10289+008 & BD+01 2447 & >1620 & {\dots} & {\dots} & 31.72 & 3 & 1620.71 & 2.36 & H$\alpha$, Ca IRT$_\mathrm{c}$, Ca IRT$_\mathrm{b}$ \\
J10564+070 & CN Leo & >815 & {\dots} & 2.70 & 2.70 & 3 & 815.72 & 6.08 & dLW (NIR), dLW (VIS), CCF-contrast (VIS) \\
J11033+359 & Lalande 21185 & >2498 & {\dots} & 53.33 & 56.15 & 17 & 2498.06 & 3.86 & NaD$_2$, H$\alpha$, NaD$_1$, NaD$_2$, dLW (VIS), CCF-contrast (VIS), He 10830 \AA, NaD$_1$, dLW (NIR), H$\alpha$, FeH WFB, H$\alpha$, CCF-contrast (VIS), CaH$_2$, NaD$_2$, NaD$_1$, CaH$_3$ \\
J11033+359 & Lalande 21185 & 536 & {\dots} & 53.33 & 56.15 & 15 & 2498.06 & 3.86 & NaD$_2$, He 10830 \AA, CaH$_3$, dLW (VIS), dLW (NIR), H$\alpha$, CCF-contrast (VIS), dLW (NIR), CCF-FWHM (VIS), NaD$_1$, He 10830 \AA, H$\alpha$, Ca IRT$_\mathrm{c}$, VO 7940 \AA, CRX (NIR) \\
J11033+359 & Lalande 21185 & 318 & {\dots} & 53.33 & 56.15 & 3 & 2498.06 & 3.86 & VO 7940 \AA, NaD$_1$, NaD$_1$ \\
J11417+427 & Ross 1003 & >1193 & {\dots} & {\dots} & 71.50 & 4 & 1193.84 & 25.16 & RVs, FeH WFB, H$\alpha$, CCF-FWHM (VIS) \\
J11509+483 & GJ 1151 & 958 & {\dots} & 150.92 & 125.00 & 6 & 2356.70 & 4.09 & CCF-FWHM (VIS), H$\alpha$, dLW (NIR), TiO 8428 \AA, Ca IRT$_\mathrm{b}$, CaH$_3$ \\
J12123+544S & HD 238090 & 535 & 547.50 & 38.42 & 96.70 & 8 & 1192.92 & 3.31 & Pa$\beta$, Ca IRT$_\mathrm{b}$, Ca IRT$_\mathrm{a}$, He 10830 \AA, Ca IRT$_\mathrm{c}$, NaD$_2$, CCF-FWHM (VIS), NaD$_1$ \\
J13299+102 & BD+11 2576 & 1099 & 547.50 & 30.78 & 30.00 & 24 & 2384.60 & 3.62 & Ca IRT$_\mathrm{b}$, Ca IRT$_\mathrm{c}$, dLW (VIS), Ca IRT$_\mathrm{a}$, NaD$_1$, CCF-contrast (VIS), CCF-FWHM (VIS), NaD$_2$, dLW (NIR), Ca IRT$_\mathrm{c}$, RVs, Ca IRT$_\mathrm{b}$, FeH WFB, Pa$\beta$, Ca IRT$_\mathrm{c}$, CRX (NIR), TiO 7048 \AA, Ca IRT$_\mathrm{c}$, dLW (VIS), Ca IRT$_\mathrm{a}$, Pa$\beta$, TiO 8858 \AA, Ca IRT$_\mathrm{b}$, VO 7940 \AA \\
J13299+102 & BD+11 2576 & 280 & 547.50 & 30.78 & 30.00 & 4 & 2384.60 & 3.62 & H$\alpha$, Ca IRT$_\mathrm{b}$, Ca IRT$_\mathrm{a}$, NaD$_2$ \\
J13457+148 & HD 119850 & 262 & {\dots} & 48.51 & 52.30 & 7 & 815.88 & 3.57 & CCF-FWHM (VIS), RVs, dLW (VIS), H$\alpha$, Ca IRT$_\mathrm{b}$, CaH$_2$, NaD$_1$ \\
J13457+148 & HD 119850 & >815 & {\dots} & 48.51 & 52.30 & 6 & 815.88 & 3.57 & H$\alpha$, CCF-contrast (VIS), dLW (NIR), He 10830 \AA, RVs, NaD$_2$ \\
J14257+236W & BD+24 2733A & 208 & 438.00 & {\dots} & 111.00 & 5 & 1290.66 & 4.02 & Pa$\beta$, dLW (NIR), Ca IRT$_\mathrm{b}$, Ca IRT$_\mathrm{c}$, Ca IRT$_\mathrm{a}$ \\
J16167+672S & HD 147379 & >1634 & {\dots} & 22.06 & 22.00 & 6 & 1634.72 & 5.84 & CCF-FWHM (VIS), Ca IRT$_\mathrm{b}$, Ca IRT$_\mathrm{c}$, Ca IRT$_\mathrm{a}$, dLW (NIR), dLW (VIS) \\
J16167+672S & HD 147379 & 203 & {\dots} & 22.06 & 22.00 & 5 & 1634.72 & 5.84 & Ca IRT$_\mathrm{b}$, CCF-contrast (VIS), Ca IRT$_\mathrm{c}$, dLW (VIS), CCF-FWHM (VIS) \\
J16167+672S & HD 147379 & 502 & {\dots} & 22.06 & 22.00 & 4 & 1634.72 & 5.84 & Pa$\beta$, Ca IRT$_\mathrm{c}$, Ca IRT$_\mathrm{b}$, H$\alpha$ \\
J16167+672S & HD 147379 & 235 & {\dots} & 22.06 & 22.00 & 3 & 1634.72 & 5.84 & VO 7940 \AA, Ca IRT$_\mathrm{a}$, CCF-FWHM (VIS) \\
J17303+055 & BD+05 3409 & >904 & {\dots} & 33.51 & 34.60 & 3 & 904.65 & 3.40 & Ca IRT$_\mathrm{c}$, Ca IRT$_\mathrm{a}$, CCF-FWHM (VIS) \\
J17578+046 & Barnard's Star & >2463 & {\dots} & 114.06 & 145.00 & 12 & 2463.51 & 2.80 & CCF-FWHM (VIS), dLW (NIR), CaH$_3$, dLW (VIS), CCF-contrast (VIS), dLW (NIR), TiO 7048 \AA, TiO 8428 \AA, Fe 8689 \AA, RVs, CRX (NIR), Ca IRT$_\mathrm{c}$ \\
J17578+046 & Barnard's Star & 751 & {\dots} & 114.06 & 145.00 & 12 & 2463.51 & 2.80 & dLW (NIR), dLW (NIR), dLW (VIS), TiO 7048 \AA, CCF-contrast (VIS), CCF-FWHM (VIS), TiO 7048 \AA, FeH WFB, CCF-FWHM (VIS), He 10830 \AA, dLW (NIR), H$\alpha$ \\
J17578+046 & Barnard's Star & 268 & {\dots} & 114.06 & 145.00 & 5 & 2463.51 & 2.80 & Fe 8689 \AA, CaH$_3$, CRX (VIS), Pa$\beta$, TiO 8428 \AA \\
J17578+046 & Barnard's Star & 459 & {\dots} & 114.06 & 145.00 & 3 & 2463.51 & 2.80 & He 10830 \AA, CCF-FWHM (VIS), dLW (NIR) \\
J18224+620 & GJ 1227 & 490 & {\dots} & {\dots} & {\dots} & 3 & 1608.73 & 3.52 & CCF-contrast (VIS), TiO 7048 \AA, dLW (VIS) \\
J18346+401 & LP 229-017 & >1759 & 1460.00 & {\dots} & 40.20 & 3 & 1759.09 & 3.31 & Ca IRT$_\mathrm{b}$, Ca IRT$_\mathrm{a}$, CCF-contrast (VIS) \\
J19169+051N & V1428 Aql & 241 & {\dots} & 50.43 & 46.00 & 4 & 621.52 & 3.11 & TiO 7048 \AA, TiO 8428 \AA, FeH WFB, He 10830 \AA \\
J19169+051N & V1428 Aql & >621 & {\dots} & 50.43 & 46.00 & 3 & 621.52 & 3.11 & CCF-FWHM (VIS), Ca IRT$_\mathrm{c}$, dLW (VIS) \\
J20260+585 & Wolf 1069 & >1453 & {\dots} & 156.30 & 57.70 & 17 & 1453.99 & 2.94 & CaH$_3$, H$\alpha$, Ca IRT$_\mathrm{b}$, TiO 7048 \AA, NaD$_1$, NaD$_2$, dLW (NIR), CCF-FWHM (VIS), TiO 7048 \AA, RVs, CCF-contrast (VIS), dLW (VIS), VO 7940 \AA, CRX (VIS), VO 7434 \AA, FeH WFB, H$\alpha$ \\
J20533+621 & HD 199305 & >1400 & {\dots} & 159.61 & {\dots} & 3 & 1400.98 & 2.85 & dLW (NIR), H$\alpha$, NaD$_1$ \\
J21164+025 & LSPM J2116+0234 & >1036 & {\dots} & 44.15 & 42.68 & 3 & 1036.03 & 5.24 & CRX (VIS), CCF-contrast (VIS), TiO 8858 \AA \\
J21466+668 & G 264-012 & >1306 & {\dots} & 15.17 & 100.00 & 4 & 1306.65 & 4.39 & Ca IRT$_\mathrm{b}$, CaH$_3$, H$\alpha$, FeH WFB \\
J22057+656 & G 264-18 A & >939 & 3650.00 & {\dots} & 120.50 & 4 & 939.69 & 3.87 & TiO 8428 \AA, Ca IRT$_\mathrm{b}$, H$\alpha$, Ca IRT$_\mathrm{a}$ \\
J22125+085 & Wolf 1014 & 1092 & {\dots} & {\dots} & {\dots} & 3 & 2298.73 & 8.10 & RVs, CaH$_3$, Ca IRT$_\mathrm{c}$ \\
J22137-176 & LP 819-052 & >781 & {\dots} & {\dots} & 116.40 & 3 & 781.82 & 7.64 & RVs, CRX (NIR), CaH$_3$ \\
J22330+093 & BD+08 4887 & 1115 & 1131.50 & 37.66 & 37.80 & 6 & 2248.89 & 3.77 & H$\alpha$, Ca IRT$_\mathrm{a}$, Ca IRT$_\mathrm{c}$, Ca IRT$_\mathrm{b}$, CCF-FWHM (VIS), CaH$_3$ \\
J22330+093 & BD+08 4887 & 320 & 1131.50 & 37.66 & 37.80 & 3 & 2248.89 & 3.77 & Ca IRT$_\mathrm{b}$, Ca IRT$_\mathrm{c}$, Ca IRT$_\mathrm{a}$ \\
J22532-142 & IL Aqr & >1261 & {\dots} & 51.39 & 81.00 & 10 & 1261.67 & 168.60 & dLW (VIS), CCF-contrast (VIS), CaH$_3$, TiO 7048 \AA, Ca IRT$_\mathrm{b}$, CaH$_3$, VO 7940 \AA, TiO 8858 \AA, CRX (VIS), Ca IRT$_\mathrm{a}$ \\
J22565+165 & HD 216899 & >2499 & 1971.00 & 39.85 & 39.50 & 21 & 2499.96 & 3.64 & dLW (VIS), CCF-contrast (VIS), Ca IRT$_\mathrm{c}$, NaD$_1$, Ca IRT$_\mathrm{a}$, H$\alpha$, dLW (NIR), Ca IRT$_\mathrm{b}$, NaD$_2$, RVs, He 10830 \AA, CCF-FWHM (VIS), CRX (NIR), He 10830 \AA, NaD$_1$, FeH WFB, Pa$\beta$, H$\alpha$, TiO 8428 \AA, Ca IRT$_\mathrm{a}$, Ca IRT$_\mathrm{b}$ \\
J22565+165 & HD 216899 & 577 & 1971.00 & 39.85 & 39.50 & 10 & 2499.96 & 3.64 & CaH$_3$, H$\alpha$, TiO 7048 \AA, Ca IRT$_\mathrm{b}$, Ca IRT$_\mathrm{c}$, Ca IRT$_\mathrm{a}$, NaD$_2$, CCF-FWHM (VIS), dLW (NIR), Pa$\beta$ \\
J22565+165 & HD 216899 & 321 & 1971.00 & 39.85 & 39.50 & 6 & 2499.96 & 3.64 & TiO 7048 \AA, RVs, dLW (NIR), CCF-contrast (VIS), VO 7940 \AA, CaH$_3$ \\
J22565+165 & HD 216899 & 252 & 1971.00 & 39.85 & 39.50 & 3 & 2499.96 & 3.64 & TiO 7048 \AA, CCF-contrast (VIS), dLW (VIS) \\
J23216+172 & LP 462-027 & 248 & {\dots} & 34.46 & 74.70 & 4 & 1186.77 & 2.98 & CCF-contrast (VIS), dLW (VIS), TiO 7048 \AA, VO 7434 \AA \\
J23419+441 & HH And & >1383 & {\dots} & 2.07 & 106.00 & 8 & 1383.13 & 2.61 & CCF-contrast (VIS), TiO 7048 \AA, H$\alpha$, dLW (VIS), VO 7940 \AA, NaD$_1$, Ca IRT$_\mathrm{b}$, Ca IRT$_\mathrm{a}$ \\
J23492+024 & BR Psc & >2498 & {\dots} & 50.83 & 49.90 & 12 & 2498.25 & 2.41 & H$\alpha$, dLW (NIR), RVs, dLW (VIS), Ca IRT$_\mathrm{c}$, NaD$_1$, CaH$_3$, TiO 7048 \AA, Ca IRT$_\mathrm{b}$, CRX (NIR), Ca IRT$_\mathrm{a}$, TiO 8858 \AA \\
            \end{longtable}
        \end{landscape}

    } 


    {\onecolumn
        \section{Clustering diagrams of the rotation periods detected in this work}
        \label{app:new_periods}

        \begin{figure}[!ht]
            \centering
            \includegraphics[width=\textwidth]{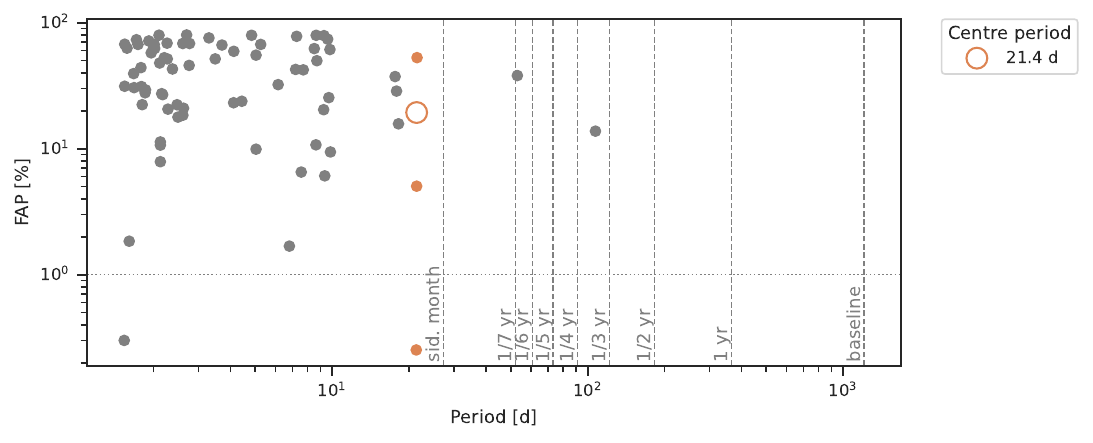}\\
            \caption{Results of the clustering algorithm for BD+09~2636 using the pre-whitening method with the yearly aliases removed. The FAP of \SI{1}{\percent} is marked by the grey horizontal dotted line and the grey vertical dashed lines show different periods of interest. The error bars of the data points correspond to the peak width in the GLS.}
            \label{fig:clustering_J12312+086}
        \end{figure}

        \begin{figure}[!ht]
            \centering
            \includegraphics[width=\textwidth]{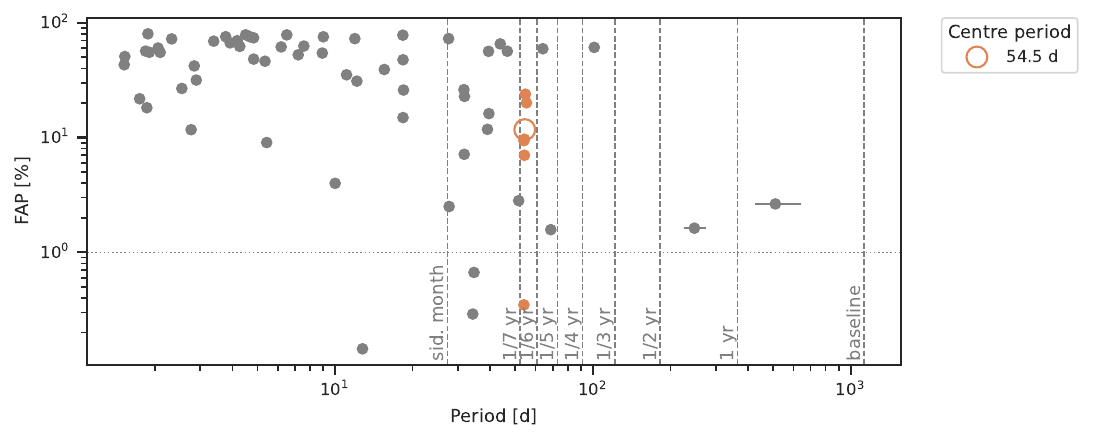}\\
            \caption{Same as Fig.~\ref{fig:clustering_J12312+086}, but for BD-07~3856.}
            \label{fig:clustering_J14307-086}
        \end{figure}

        \begin{figure}[!ht]
            \centering
            \includegraphics[width=\textwidth]{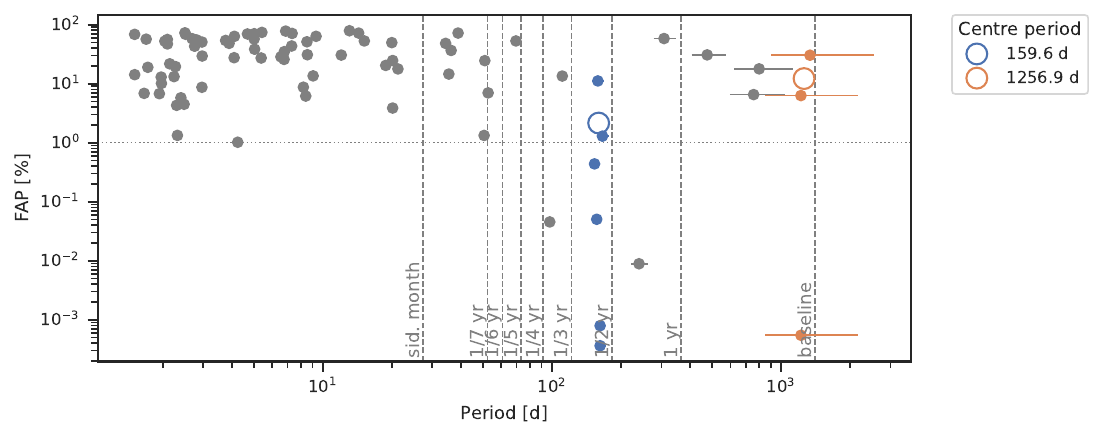}\\
            \caption{Same as Fig.~\ref{fig:clustering_J12312+086}, but for HD~199305.}            
            \label{fig:clustering_J20533+621}
        \end{figure}

    } 

\end{appendix}

\end{document}